\documentclass{article}
\usepackage[a4paper, total={6in, 9in}]{geometry} 
\usepackage{graphicx}
\usepackage{caption}
\usepackage{subcaption}
\usepackage{mathrsfs}
\usepackage{amsmath}
\usepackage{multirow}
\usepackage[table,xcdraw]{xcolor}
\usepackage[super,sort&compress]{natbib}
\bibpunct{}{}{,}{s}{,}{,}
\makeatletter
\renewcommand\@biblabel[1]{#1.}
\makeatother

\let\chapter\section
\usepackage[ruled,vlined]{algorithm2e} 

\usepackage{setspace}
\renewcommand\eqref[1]  {Eq.\ (\ref{#1})}
\newcommand\eqsref[1] {Eqs.\ (\ref{#1})}

\title{
Rapid multi-component phase-split calculations using volume functions and reduction methods
}

\author{
Mohamad Fathi\thanks{email: m.fathi@tudelft.nl ; orcid: https://orcid.org/0000-0001-9122-228X}  
and 
Stefan Hickel\thanks{email: s.hickel@tudelft.nl ; orcid: https://orcid.org/0000-0002-7463-9531}
}

\date{
Aerodynamics Group, Faculty of Aerospace Engineering, Delft University of Technology\\
Kluyverweg 2, 2629 HS Delft, The Netherlands
}
\begin{document}
\maketitle

\begin{abstract}
We present a new family of fast and robust methods for the calculation of the vapor-liquid equilibrium at isobaric-isothermal (PT-flash), isochoric-isothermal (VT-flash), isenthalpic-isobaric (HP-flash), and isoenergetic-isochoric (UV-flash) conditions. The framework is provided by formulating phase-equilibrium conditions for multi-component mixtures in an effectively reduced space based on the molar specific value of the recently introduced volume function derived from the Helmholtz free energy. The proposed algorithmic implementation can fully exploit the optimum quadratic convergence of a Newton method with the analytical Jacobian matrix. This paper provides all required exact analytic expressions for the general cubic equation of state. Computational results demonstrate the effectivity and efficiency of the new methods. Compared to conventional methods, the proposed reduced-space iteration leads to a considerable speed-up as well as to improved robustness and better convergence behavior near the spinodal and coexistence curves of multi-component mixtures, where the preconditioning by the reduction method is most effective. 
\end{abstract}


\section{Introduction}

Robust, computationally efficient and accurate phase splitting or flash calculations play a crucial role in many engineering disciplines, such as chemical-process and reservoir simulations. In Computational Fluid Dynamics (CFD) simulations of realistic multi-component vapor-liquid fluid flows, millions of phase equilibrium calculations are required every time step in the form of either the VT-flash or UV-flash, depending on the chosen formulation of the governing equations: The VT-flash is needed in cases where the overall specific volume, temperature and composition are known, such as for the carbon dioxide injection into subsurface reservoirs \citep{rongy2012mixing,POLIVKA2014149}. Methods that solve the compressible Navier-Stokes equations based on the conservation laws for mass, linear momentum and total energy, such as applied for the simulation of the trans-critical vaporization of liquid fuels \citep{matheis2018multi,MA2019231,Rodriguez:2019fi,Tudisco:2019dw}, require a UV-flash, where the input is the overall specific internal energy, volume and composition. The calculation of thermodynamic equilibrium properties of multi-component multi-phase mixtures typically consumes more than three quarters of the total computational time \citep{abhvani1987development,MICHELSEN19981} and thus imposes severe limitation on the tractable space-time resolution or even the computational feasibility of such numerical simulations. At the same time, flash algorithms for CFD applications have to be fault tolerant and robust, because even a method that fails to converge only once in a billion will eventually spoil the entire simulation.  

The simplest case and workhorse of most phase-equilibrium calculations is the so-called PT-flash, where the equilibrium pressure and temperature of the mixture are already given. Most methods for calculating the isobaric-isothermal equilibrium volume fractions and compositions follow the approach proposed by Michelsen \citep{michelsen1982stability,michelsen1982isothermal}.
For solving flash problems at conditions other than constant pressure and temperature, \citet{michelsen1999state} introduced an indirect method based on nesting simple and robust PT-flash calculations. For VT-flashes, for example, Michelsen's method aims to find the pressure at which the PT-flash results in the given total specific volume. This results in an optimization algorithm, in which the pressure is adjusted in the outer loop and a PT-flash is solved in the inner loop. Accordingly, UV-flashes are solved by a bi-variant optimization of temperature and pressure corresponding to the given internal energy and volume, which define the thermodynamic state, for example, in mass and energy conservative Navier-Stokes solvers \cite{matheis2018multi}.

Nested algorithms based on the PT-flash are also attractive for mixtures with many components, because they offer the possibility of adopting reduction methods \citep{michelsen1986simplified}, which provide a considerable speedup and, in addition, improve the robustness of the algorithm  \citep{petitfrere2015comparison}. The first reduction method was introduced by \citet{michelsen1986simplified}, who found that the phase-splitting problem is fully defined by only three reduced parameters regardless of the number of components when all Binary Interaction Coefficients (BICs) are zero. \citet{hendriks1992application} successfully generalized the method for cases with some non-zero BICs through an eigenvalue analysis of the binary interaction matrix. \citet{nichita2011new} found a new set of reduced parameters for PT-flash calculations, for which they demonstrated a notable decrease in the number of iterations relative to previous reduction methods specifically near the phase boundary and the critical point.

Employing a direct VT-flash, on the other hand, could considerably reduce the computational time by eliminating the outer pressure iteration loop, provided that the method itself would be fast and robust enough. To this end, \citet{mikyvska2011new} introduced an alternative formulation of the VT-flash problem based on a new thermodynamic function, the so-called volume function. They solved the problem directly by a successive-substitution iteration (SSI) algorithm with nearly the same number of iterations as a conventional PT-flash based solver requires for one inner iteration loop. Recently, \citet{jindrova2015general} and \citet{nichita2018new} presented methods for solving the VT-flash problem via direct minimization of the total Helmholtz free energy. \citet{cismondi2018new} directly included the pressure equality and volume constraint in a new algorithm very similar to the PT-flash, and showed about $20\%$ reduction in the computational time compared to Michelsen's nested optimization technique. However, for working fluids with a large number of components, these methods lead to a significantly stronger increase of the computational time compared to the nested approach that benefits from the quadratic Newton-Raphson convergence rate in the reduced space.

Extending the work of \citet{mikyvska2011new} and \citet{nichita2011new}, we present a very fast and robust method for direct vapor-liquid phase-split calculations based on formulating phase equilibrium conditions in terms of the molar specific value of Miky{\v{s}}ka and Firoozabadi's volume function (instead of fugacity coefficients) and a corresponding reduction method. This new formulation allows us to solve isothermal flashes (both PT and VT) directly and with the exact analytical Jacobian matrix, which results in optimum quadratic convergence of the Newton-Raphson method.  Non-isothermal cases, such as UV and HP flashes, are solved through nested univariate optimization with the corresponding isothermal flash (PT for HP and VT for UV) and the readily available specific heat capacity at constant pressure (for HP-flash) or at constant volume (for UV-flash) as exact Jacobian. 

This paper is structured as follows: First, the mathematical description of the equilibrium problem is reformulated based on the molar specific values of the volume function for the vapor and liquid phases. Next, the classical reduction method is presented along with the derivation of new reduced parameters in the context of the new formulations, and all other thermodynamic relations required for non-isothermal flashes are also derived from the reduced parameters. Then, algorithms based on the Newton-Raphson method with the analytical Jacobian matrix for the direct solution of isothermal flashes and for the indirect solution of non-isothermal ones, are presented. Last but not least, the reliability and efficiency of proposed algorithms and its significantly improved computational performance compared to a recently published implementation for high-fidelity CFD simulations \citep{matheis2018multi} will be demonstrated and discussed for different multi-component mixtures at various thermodynamic conditions.

\section{Thermodynamic equilibrium formulation}

According to the Gibbsian thermodynamics \cite{firoozabadi2015thermodynamics}, a multi-component system consisting of vapor and liquid phases is in  equilibrium when the temperatures, pressures, and chemical potentials of phases are equal, that is, 
\begin{equation}
{T^L} = {T^V},\quad {p^L} = {p^V},\quad \mu _i^L = \mu _i^V,
\end{equation}
where $T$, $p$, and $\mu_i$ are temperature, pressure, and chemical potential of component $i=\lbrace1 \ldots n \rbrace$ in a mixture with $n$ components, and superscripts $L$ and $V$ refer to values of the liquid and vapor phases.

The pressures can be computed as a function of temperature, molar specific volume and composition of each phase using the general form of the cubic equation of states (EoS) 
\begin{equation}\label{GCEoS}
p = \frac{{RT}}{{v - b}} - \frac{a}{{(v + {\delta _1}b)(v + {\delta _2}b)}} \ ,
\end{equation}
where $\delta_1$ and $\delta_2$ are the two EoS parameters (see below), $R$ is the universal gas constant, and $v$ is the molar specific volume of the mixture. The energy and co-volume parameters $a$ and $b$ are usually computed using the van der Waals mixing rules
\begin{equation}\label{eq:adef}
a = \sum\limits_{i = 1}^n {\sum\limits_{j = 1}^n {{z_i}{z_j}\left( {1 - {\kappa _{ij}}} \right)\sqrt {{\hat{a}_i}{\hat{a}_j}} } },
\end{equation}
and 
\begin{equation}\label{eq:covolume}
b = \sum\limits_{i = 1}^n {{z_i}{\hat{b}_i}} \ ,
\end{equation}
in which $z_i$ is the mole fraction of the component $i$, and $\kappa _{ij}$ is the binary interaction coefficient between component $i$ and $j$ in the mixture. $\hat{a}_i$ and $\hat{b}_i$ are energy and co-volume parameters of the pure component $i$, which are obtained through
\begin{equation}
{\hat{a}_i} = {\Omega _a}\frac{{{R^2}T_{ci}^2}}{{{p_{ci}}}}{\left[ {1 + c({\omega _i})\left( {1 - \sqrt { \frac{T}{T_{ci}} } }\right)} \right]^2} \\
\end{equation}
and 
\begin{equation}
{\hat{b}_i} = {\Omega _b}\frac{{R{T_{ci}}}}{{{p_{ci}}}} \ ,
\end{equation}
where $T_{ci}$, $p_{ci}$ are critical temperature and pressure of the component $i$. The two constants ${\Omega _a}$ and ${\Omega _b}$ as well as the form of the function of ${c({\omega _i})}$, in which ${\omega _i}$ is the acentric factor, depend of the selected cubic EoS: for instance, in the Peng-Robinson (PR) EoS: $\delta_1=1+\sqrt{2}$ and $\delta_2=1-\sqrt{2}$, which result in ${\Omega _a=0.45724}$, ${\Omega _b=0.0778}$, and the functional ${c({\omega _i})}$ is:
$${c({\omega _i})} = \left\{ {\begin{array}{*{20}{c}}
{0.37464 + 1.54226{\omega _i} - 0.26992\omega _i^2}&\mathrm{~~ , for~~ }{{\omega _i} < 0.5}\\
{0.3796 + 1.485{\omega _i} - 0.1644\omega _i^2 + 0.01667\omega _i^3}&\mathrm{~~ , for~~ }{{\omega _i} \ge 0.5} 
\end{array}} \right. \ .$$ 
In the Soave-Redlich-Kwong (SRK) EoS, with $\delta_1=0$ and $\delta_2=1$, they are ${\Omega _a=0.42748}$, ${\Omega _b=0.08664}$, and 
${c({\omega _i})}={0.48508 + 1.55171{\omega _i} - 0.15613\omega _i^2}.$

The equality of chemical potentials is typically expressed in terms of the K-factor (also named K-value or equilibrium ratio, which is the ratio of the mole fractions in the vapor ($y$) and liquid ($x$) phases), and the fugacity coefficient derived from the Gibbs free energy. The logarithmic form of this relation for a two-phase vapor-liquid mixture is
\begin{equation}\label{eq:fugacityK}
\ln {K_i} = \ln \varphi _i^L - \ln \varphi _i^V \mathrm{~~ , for~~ }  i = 1 \ldots n \ ,
\end{equation}
with $\varphi$ as the fugacity coefficient and ${K}$ as the K-factor. \citet{mikyvska2011new} derived a new thermodynamic function for the evaluation of the equilibrium ratio via minimization of the Helmholtz free energy that uses the specific volume, temperature, and mole fractions as its primary variables and eliminates the need for knowing the equilibrium pressure and for solving the state equation for the stable volume. They proved that the following relationship exists between the K-factor and the volume function coefficient for the liquid and vapor phases: 
\begin{equation}
{K_i} = \frac{{{v^V}\Phi_i ({v^V},T,{y_1},...,{y_n})}}{{{v^L}\Phi_i ({v^L},T,{x_1},...,{x_n})}} \ ,
\end{equation}
in which $\Phi_i$ is the volume function coefficient of the component $i$ and can be computed analytically as a function of temperature, specific volume, and mole fractions via 
\begin{equation}\label{eq:volfundef}
\ln {\Phi _i} = \int\limits_v^{ + \infty } {\left[ {\frac{1}{v} - \frac{1}{{RT}}{{\left( {\frac{{\partial p}}{{\partial {z_i}}}} \right)}_{T,v,{z _{j \ne i}}}}} \right]dv}.
\end{equation}
We define 
\begin{equation}\label{eq:vulacity}
{\psi _i} \equiv {v}{\Phi _i}
\end{equation}
as the molar specific value of the volume function, such that, instead of using \eqref{eq:fugacityK}, the natural logarithm of K-factors can be calculated by
\begin{equation}\label{eq:vulacityK}
\ln {K_i} = \ln \psi _i^V - \ln \psi _i^L\mathrm{~~ , for~~ } i = 1 \ldots n \ .
\end{equation}
It can be shown that this (molar) specific volume function is related to the fugacity coefficient via ${\psi _i} = RT/p{\varphi _i}$. By substituting  the general cubic EoS (\ref{GCEoS}) for the evaluation of the partial pressure term in the integral (\ref{eq:volfundef}), the following expression is obtained for this new thermodynamic function:
\begin{equation}\label{eq:modvolfun}
\ln {\psi _i} = \ln \left( {v - b} \right) - \frac{{{{\hat b}_i}}}{{v - b}} + \frac{{av{{\hat b}_i}/\left( {bRT} \right)}}{{\left( {v + {\delta _1}b} \right)\left( {v + {\delta _2}b} \right)}} - \frac{{a{{\hat b}_i} - 2b{g_i}}}{{({\delta _1} - {\delta _2}){b^2}RT}}\ln \left( {\frac{{v + {\delta _1}b}}{{v + {\delta _2}b}}} \right)
\end{equation}
where $g_i$ is 
\begin{equation}\label{eq:gidef}
g_i=\sum\limits_{j = 1}^n {z_j\left( {1 - {\kappa _{ij}}} \right)\sqrt {{\hat{a}_i}{\hat{a}_j}} }  \mathrm{~~ , for~~ }  i = 1 \ldots n .
\end{equation}

The equality of chemical potentials and component material balances can be systematically expressed by means of K-factors in such a way that the vapor mole fraction $\theta$ is determined by the classic Rachford-Rice equation  
\begin{equation}\label{eq:rachford}
\sum\limits_{i = 1}^n {\frac{{{{\hat z}_i}({K_i} - 1)}}{{1 + \theta ({K_i} - 1)}}}  = 0 \ ,
\end{equation}
where ${\hat z}_i$ is the overall mole fractions of component $i$ in the feed. Then, molar compositions of the liquid and vapor phases are obtained:
\begin{equation}\label{eq:composition}
{x_i} = \frac{{{{\hat z}_i}}}{{1 + \theta \left( {{K_i} - 1} \right)}} \mathrm{\quad and \quad} {y_i} = {x_i}{K_i}  \mathrm{\quad, for~~ }  i = 1 \ldots n .
\end{equation}
This formulation leads to a remarkable reduction in the number of variables in isothermal flash calculations; we know the overall composition of the feed hence, by knowing the $n$ K-factors, we can compute the molar compositions of the vapor and liquid. The molar specific volumes of phases can then be computed by evaluating the state equations separately for vapor and liquid based on the given pressure in PT-flash calculations \citep{michelsen2007thermodynamic}, or by solving the pressure equality equation along with a volume constraint based on the given volume in VT-flash calculations \citep{cismondi2018new}.

\section{Reduction method}

The basic idea of all reduction methods is to calculate the K-factors in a lower-dimensional hyperspace spanned by parameters that are independent of  the number of components in the mixture. 
According to the classical theory of reduction \citep{firoozabadi2015thermodynamics}, such reduced parameters can be obtained by decomposing the symmetric matrix $\beta_{ij} = 1 - {\kappa_{ij}}$ that represents the binary interactions into matrices composed of its eigenvectors and eigenvalues, that is, 
\begin{equation}\label{eq:beta1}
{\beta} = SD{S^{ - 1}} = SD{S^T},
\end{equation}
in which the diagonal matrix ${D} = {\mathop{\rm diag}\nolimits} \left( {{\lambda _1},...,\;{\lambda _n}} \right)$ represents the eigenvalues $\lambda_{i}(i= 1 \ldots n)$ of the matrix $\beta$, and the orthogonal matrix $S = ({{\vec s}_1},...,{{\vec s}_n})$ includes the corresponding eigenvectors ${\vec s}_i(i= 1 \ldots n)= ({s_{i1}},...,{s_{in}})^{T}$. 
For most mixtures with a large number of components, only a few $(m<n)$ eigenvalues are significant as a result of negligible binary interactions between many components; we can hence use the following approximation for the evaluation of the entries of the matrix $\beta$:
\begin{equation}\label{beta2}
{\beta _{ij}} = \sum\limits_{k = 1}^n {{\lambda _k}{s_{ki}}{s_{kj}}}  \approx \sum\limits_{k = 1}^m {{\lambda _k}{s_{ki}}{s_{kj}}}.
\end{equation}
Defining ${\hat s_{ki}} \equiv {s_{ki}}\sqrt {{{\hat a}_i}}$ as entries of the reduction matrix with size of $m \times n$, we can express $g_i(i= 1 \ldots n)$ in \eqref{eq:gidef} as 
\begin{equation}\label{eq:gireducedspace}
{g_i} = \sum\limits_{j = 1}^n {{z_j}(\sum\limits_{k = 1}^m {{\lambda _k}} {s_{ki}}{s_{kj}})\sqrt {{{\hat a}_i}{{\hat a}_j}} }  = \sum\limits_{k = 1}^m {{\lambda _k}{{\hat s}_{ki}}(\sum\limits_{j = 1}^n {{z_j}{{\hat s}_{kj}}} ) = } \sum\limits_{k = 1}^m {{\lambda _k}{{\hat s}_{ki}}{q_k}},
\end{equation}
as a function of the reduced parameters  
\begin{equation}\label{eq:reducedparameters}
{q_k} = \sum\limits_{i = 1}^n {{z_i}{{\hat s}_{ki}}} \mathrm{~~ , for~~ }  k = 1 \ldots m  \ .
\end{equation}
Similarly, the energy parameter $a$ of the mixture in \eqref{eq:adef} can be calculated from these reduced parameters via
\begin{equation}\label{eq:areducedspace}
a = \sum\limits_{i = 1}^n {\sum\limits_{j = 1}^n {{z_i}{z_j}(\sum\limits_{k = 1}^m {{\lambda _k}} {s_{ki}}{s_{kj}})\sqrt {{{\hat a}_i}{{\hat a}_j}} } }  = \sum\limits_{k = 1}^m {{\lambda _k}{{(\sum\limits_{i = 1}^n {{z_i}{{\hat s}_{ki}}} )^2}}}  = \sum\limits_{k = 1}^m {{\lambda _k}q_k^2}.
\end{equation}
Then, an equation for evaluation of the molar specific value of the volume functions, $\psi _i$, can be derived by substituting $g_i(i= 1 \ldots n)$ and $a$ into \eqref{eq:modvolfun} using \eqsref{eq:gireducedspace} and (\ref{eq:areducedspace}):
\begin{equation}\label{vulacity2}
\ln {\psi _i} = \sum\limits_{k = 1}^m {{h_k}{{\hat s}_{ki}} + {h_{m + 1}}{\hat{b}_i} + {h_{m + 2}}}  \mathrm{~~ , for~~ }  i = 1 \ldots n ,
\end{equation}
where coefficients $h$ are functions of $q_k(k= 1 \ldots m)$, $b$, and $v$:
\begin{equation}
\label{eq:hcoefficients}
\begin{aligned}
{h_k} &= 2{\lambda _k}{q_k}\ln \left[ {(v + {\delta _1}b)/(v + {\delta _2}b)} \right]/\left[ {({\delta _1} - {\delta _2})bRT} \right]  \mathrm{~~ , for~~ }  k = 1 \ldots m 
\\
{h_{m + 1}} &= \sum\limits_{k = 1}^m {{\lambda _k}q_k^2\left\{ \begin{array}{l}
vb/\left[ {(v + {\delta _1}b)(v + {\delta _2}b)} \right] -  \\ 
\ln \left[ {(v + {\delta _1}b)/(v + {\delta _2}b)} \right]/({\delta _1} - {\delta _2}) \\ \end{array} \right\}/\left( {RTb^2} \right)}  - 1/{\left( {v - b} \right)},
\\
{h_{m + 2}} &= \ln (v - b).
\end{aligned}
\end{equation}
Because the entries of reduction matrix ${\hat s}_{ki}$ and ${\hat b}_i$ are equal in the liquid and vapor phases, all K-factors can be computed from:
\begin{equation}\label{eq:Kfactors}
\ln {K_i} = \sum\limits_{k = 1}^m {h_k^\Delta {{\hat s}_{ki}} + h_{m + 1}^\Delta {\hat{b}_i} + h_{m + 2}^\Delta } \ ,
\end{equation}
with $h_{\alpha}^\Delta  \equiv h_{\alpha}^V - h_{\alpha}^L\; ({\alpha}=1,...,m+2)$. Performing the calculations in the $h$-space with size $m+2$ leads to another remarkable reduction in the number of variables in the multi-component flash calculation, i.e. to dimension $m+2$ instead of $n$ regardless of the number of components in the mixture.

We note that these $h$-based reduced parameters are Lagrange multipliers of the classical reduced parameters, similar to the reduced parameters introduced by \citet{nichita2011new}. Hence, the reduced-space iteration has a better condition number and will converge faster than other methods \citep{nichita2013phase}.

\subsection{Thermodynamic relations for non-isothermal flashes}\label{sec:derivatives}

For non-isothermal flash calculations, it is necessary to compute additional thermodynamic quantities such as the specific molar enthalpy, internal energy, and heat capacities at constant volume and pressure. They are typically calculated as a summation of the ideal part, which is here evaluated as a function of temperature using the 9-coefficient NASA polynomials \cite{goos2009third}, and the excess part obtained from the state equation using the reduced parameters. Overall mixture quantities are computed through
\begin{equation}
{\eta ^{mix}} = \left( {1 - \theta } \right){\eta ^L} + \theta {\eta ^V} \ ,
\end{equation}
where $\eta  \in \left\{ {u,h,{c_v},{c_p}} \right\}$ are specific internal energy, enthalpy, and heat capacities at constant volume and pressure. The molar specific internal energy of the liquid or vapor (superscripts $L$ and $V$ are not repeated for brevity) is computed via
\begin{equation}
u = \sum\limits_{i = 1}^n {{z_i}u_i^{ig}(T)}  + \frac{{a - T\left( {{{\partial a} \mathord{\left/
 {\vphantom {{\partial a} {\partial T}}} \right.
 \kern-\nulldelimiterspace} {\partial T}}} \right)}}{{\left( {{\delta _2} - {\delta _1}} \right)b}}\ln \left( {\frac{{v + {\delta _1}b}}{{v + {\delta _2}b}}} \right) \ ,
 \end{equation}
where $u_i^{ig}$ is the ideal gas (NASA polynomial) molar specific internal energy of pure component $i$; $a$ is obtained from \eqref{eq:areducedspace} and its first temperature derivative is
\begin{equation}
\frac{{\partial a}}{{\partial T}} = 2\sum\limits_{k = 1}^m {{\lambda _k}{q_k}\frac{{\partial {q_k}}}{{\partial T}}} ,
\end{equation}
with
\begin{equation}
\frac{{\partial {q_k}}}{{\partial T}} =  - \frac{R}{2}\sqrt {\frac{{{\Omega _a}}}{T}} \sum\limits_{i = 1}^n {\frac{{{z_i}c({\omega _i}){s_{ki}}}}{{{\mathop{\rm sgn}} ({\vartheta _i})}}\sqrt {\frac{{{T_{ci}}}}{{{p_{ci}}}}} } \ ,
\end{equation}
where ${\mathop{\rm sgn}} (\vartheta_i )$ is the sign function of variable $\vartheta_i  = 1 + c({\omega _i})(1 - \sqrt {T/{T_{ci}}} )$ and its value is equal to plus one for $\vartheta_i >0$ and equal to minus one otherwise. The molar specific enthalpy of the mixture is defined as
\begin{equation}
h = u + pv \ ,
\end{equation}
where $p$ is either known or computed via \eqref{GCEoS}. The molar specific heat capacity at constant volume for a multi-component mixture can be computed via 
\begin{equation}
{c_v} = \sum\limits_{i = 1}^n {{z_i}c_{v,i}^{ig}(T)}  + \frac{{T\left( {{{{\partial ^2}a} \mathord{\left/
 {\vphantom {{{\partial ^2}a} {\partial {T^2}}}} \right.
 \kern-\nulldelimiterspace} {\partial {T^2}}}} \right)}}{{\left( {{\delta _1} - {\delta _2}} \right)b}}\ln \left( {\frac{{v + {\delta _1}b}}{{v + {\delta _2}b}}} \right) \ .
\end{equation}
Here, $c_{v,i}^{ig}$ is the ideal gas molar specific heat capacity at constant volume for the component $i$, which is computed as a function of temperature using NASA polynomials, and the second derivative $\partial^2 a/\partial T^2$ is 
\begin{equation}
\frac{{{\partial ^2}a}}{{\partial {T^2}}} = 2\sum\limits_{k = 1}^m {{\lambda _k}\left[ {{{\left( {\frac{{\partial {q_k}}}{{\partial T}}} \right)}^2} + {q_k}\frac{{{\partial ^2}{q_k}}}{{\partial {T^2}}}} \right]} 
\end{equation}
with
\begin{equation}
\frac{{{\partial ^2}{q_k}}}{{\partial {T^2}}} = \frac{R}{{4T}}\sqrt {\frac{{{\Omega _a}}}{T}} \sum\limits_{i = 1}^n {\frac{{{z_i}c({\omega _i}){s_{ki}}}}{{{\mathop{\rm sgn}} ({\vartheta _i})}}\sqrt {\frac{{{T_{ci}}}}{{{p_{ci}}}}} }  \ .
\end{equation}
The molar specific heat capacity at constant pressure of the mixture is computed from the thermodynamic relation 
\begin{equation}
{c_p} = {c_v} - T\frac{{{{\left( {{{\partial p} \mathord{\left/
 {\vphantom {{\partial p} {\partial T}}} \right.
 \kern-\nulldelimiterspace} {\partial T}}} \right)}^2}}}{{{{\partial p} \mathord{\left/
 {\vphantom {{\partial p} {\partial v}}} \right.
 \kern-\nulldelimiterspace} {\partial v}}}}
 \ ,
\end{equation}
where the derivatives of pressure with respect to the specific volume and temperature are 
\begin{equation}\label{eq:dpdv}
\frac{{\partial p}}{{\partial v}} =  - \frac{{RT}}{{{{\left( {v - b} \right)}^2}}} + \frac{{a\left[ {2v + \left( {{\delta _1} + {\delta _2}} \right)b} \right]}}{{{{\left( {v + {\delta _1}b} \right)}^2}{{\left( {v + {\delta _2}b} \right)}^2}}}
\end{equation}
and
\begin{equation}
\frac{{\partial p}}{{\partial T}} = \frac{R}{{v - b}} - \frac{{{{\partial a} \mathord{\left/
 {\vphantom {{\partial a} {\partial T}}} \right.
 \kern-\nulldelimiterspace} {\partial T}}}}{{\left( {v + {\delta _1}b} \right)\left( {v + {\delta _2}b} \right)}}
 \ .
\end{equation}

\section{Numerical algorithms}

\subsection{Isothermal flashes}

\begin{algorithm}[!tb]
\SetAlgoLined
\KwResult{K-factors of a multi-component vapor-liquid equilibrium}
\vspace{2mm}
Step 0: Estimate initial values of K-factors using the input values or via the Wilson's correlation in case of blind flashes\;
\While{ convergence criteria not met }{
Step 1: Calculate $\theta$ by solving the Rachford-Rice equation (Eq.~\ref{eq:rachford})\;
\vspace{2mm}
Step 2: Determine molar compositions $x$ and $y$ (Eq.~\ref{eq:composition}) and then compute parameters $q_k$ and $b$ for both phases (Eqs.~\ref{eq:reducedparameters} and \ref{eq:covolume})\;
\vspace{2mm}
Step 3: Compute molar specific volumes $v^L$ and $v^V$ using pressure equality and volume constraint equations in case of VT-flash and two state equations of liquid and vapor in case of PT-flash\;
\vspace{2mm}
Step 4: Evaluate Jacobian matrix and update the reduced principal variables for the Newton-Raphson iteration or update the principal variables via their definitions in case of the successive substitution method\;
\vspace{2mm}
Step 5: Update K-factors (Eq.~\ref{eq:Kfactors}) and check the convergence criteria.}
\caption{VT and PT flash calculations}
\label{alg:1}
\end{algorithm}

In this section, numerical solution procedures for two important isothermal phase splitting cases, PT and VT-flashes, are presented. 
In Algorithm \ref{alg:1}, we need to estimate the initial K-factors at Step 0: if there is no promising data available (blind flash), Wilson's correlation  
\begin{equation}\label{Wilson}
K_i = \frac{{{p_{ci}}}}{p}\exp \left[ {5.373\left( {1 + {\omega _i}} \right)\left( {1 - \frac{{{T_{ci}}}}{T}} \right)} \right] \mathrm{~~ , for~~ }  i= 1 \ldots n
\end{equation}
is commonly employed for the initialization of the iteration. This is straightforward if the pressure and temperature are known as in the PT-flash; in the case of a blind VT-flash, however, the pressure is unknown. In this case, one could estimate the pressure from the state equation of the mixture by using the  total specific volume $\hat v$, temperature $T$, and overall mole fractions $\hat z_i$ as an input, but this will result in negative pressures in many cases. A simple remedy is to set a minimum value in pressure estimation \citep{cismondi2018new}, or to employ the initialization method based on the vapor pressures of the components \citep{mikyvska2012investigation,nichita2017fast}. We propose to use the geometric average of the pressures of the dew and bubble points estimated as 
\begin{equation}\label{eq:EstimPress}
p \approx \sqrt {\sum\limits_{i = 1}^n {{{\hat z}_i}p_i^{sat}} /\sum\limits_{i = 1}^n {\frac{{{{\hat z}_i}}}{{p_i^{sat}}}} } \ ,
\end{equation}
where $p_i^{sat}$ is the vapor pressure of the pure component $i$, which can be estimated from Raoult's law and Wilson's correlation. 

In Step 1, we need to solve the Rachford-Rice \eqref{eq:rachford} to determine the vapor mole fraction. 
Usually, a Newton method is coupled with a bisection method for reasons explained by \citet{michelsen2007thermodynamic}. 
In order to preserve the fully quadratic convergence rate of the Newton method, we rather use the convex transformation technique of \citet{nichita2013rapid}. In the convex transformation technique, the first and last index are assigned to the components with maximum and minimum K-factors, and vectors ${c_i} = 1/(1 - {K_i})$ and ${d_i} = ({c_1} - {c_i})/({c_n} - {c_1})$ are obtained for all components. Two convex functions of the variable $\sigma = (\theta  - {c_1})/({c_n} - \theta )$, can be then computed:
\begin{subequations}
\begin{align}
\begin{split}
G(\sigma) &= (1 + \sigma)S(\sigma )\ , 
\end{split}\\
\begin{split}
H(\sigma ) &=  - \sigma (1 + \sigma )S(\sigma )\ ,
\end{split}
\end{align}
\end{subequations}
where
\begin{equation}
S(\sigma) = \sum\limits_{i = 1}^n {\frac{{{z_i}}}{{{d_i} + \sigma(1 + {d_i})}}}.
\end{equation}
For any starting value $\sigma_0$ in the range of $\left( 0, + \infty  \right)$, monotonic convergence of the Newton iteration is guaranteed for one of these two functions. The estimated value of $\sigma$ is updated via
\begin{equation}
{\sigma _{new}} = \left\{ {\begin{array}{*{20}{c}}
{\sigma  - G(\sigma )/G'(\sigma )}&\mathrm{, for \, }{G(\sigma ) > 0}\\
{\sigma  - H(\sigma )/H'(\sigma )}&\mathrm{, for \, }{G(\sigma ) < 0}
\end{array}} \right.
\end{equation}
where $G'$ and $H'$ are derivatives of $G$ and $H$ with respect to $\sigma$:
\begin{subequations}
\begin{align}
\begin{split}
G'(\sigma) = S(\sigma ) + (1 + \sigma)S'(\sigma )\ , 
\end{split}\\
\begin{split}
H'(\sigma) =  - (1 + 2\sigma)S(\sigma) - \sigma(1 + \sigma)S'(\sigma)\ , 
\end{split}
\end{align}
\end{subequations}
and 
\begin{equation}
S'(\sigma) = \sum\limits_{i = 1}^n {\frac{{ - {z_i}(1 + {d_i})}}{{{{\left( {{d_i} + \sigma(1 + {d_i})} \right)}^2}}}} \ .
\end{equation}
The Newton iteration is repeated with $\sigma_{new}$ until the convergence criteria is met. The vapor mole fraction is then obtained via $\theta  = ({c_1} + \sigma{c_n})/(1 + \sigma).$

In Step 2, molar compositions of the liquid $x_i$ and vapor $y_i$ for $(i= 1 \ldots n)$ are computed using \eqref{eq:composition}. Then $m$ reduced parameters $q_k$ and mixture co-volume parameter $b$ are obtained using \eqsref{eq:reducedparameters} and (\ref{eq:covolume} )for both phases.

In Step 3, the energy parameter $a$ is computed for the liquid and vapor phases using their $m$ reduced parameters via \eqref{eq:areducedspace}. For the case of PT-flash, in which the value of the equilibrium pressure $p$ is given, the specific volume is then computed for the vapor and liquid phases separately. For general cubic EoS  \eqref{GCEoS} this means to find the roots of the cubic equation that is written below for the liquid phase:
\begin{equation}\label{eq:cubicvL}
v_L^3 + {\varrho _2}v_L^2 + {\varrho _1}{v_L} + {\varrho _0} = 0 \ ,
\end{equation}
where
\begin{equation}
\begin{aligned}
 {\varrho _0} &=  - {a_L}{b_L}/p - \left( {{b_L} + RT/p} \right){{\delta _1}{\delta _2}} b_L^2 \ , \\ 
 {\varrho _1} &= {{\delta _1}{\delta _2}} b_L^2 + {a_L}/p - \left( {{\delta _1} + {\delta _2}} \right) {b_L}\left( {{b_L} + RT/p} \right) \ ,\\ 
 {\varrho _2} &= \left( {{\delta _1} + {\delta _2} - 1} \right){b_L} - RT/p \ . \\ 
 \end{aligned}
 \end{equation}
The same equation is holds for the vapor phase. We use Cardino's algorithm to determine all roots of \eqref{eq:cubicvL}. If more than one real root is found, the root associated with the lowest Gibbs free energy is selected \cite{michelsen2007thermodynamic}.

For the VT-flash, in which the value of the total molar specific volume $\hat v$ is given, we first compute the molar specific volume of one phase from the volume constraint $(1 - \theta ){v_L} + \theta {v_V} = \hat v$ and then substitute it into the pressure equality equation. The resulting equation is a quintic function of the other phase specific volume that is given below for the liquid phase:
\begin{equation}\label{eq:quinticvL}
{{\varsigma _5}v_L^5 + {\varsigma _4}v_L^4 + {\varsigma _3}v_L^3 + {\varsigma _2}v_L^2 + {\varsigma _1}{v_L} + {\varsigma _0} = 0}
\end{equation}
where
\begin{equation}
\begin{aligned}
{\varsigma _0} &= (\alpha _2^L\alpha _5^V - \alpha _2^V\alpha _5^L){\theta ^3} - (\alpha _5^L\alpha _1^V - \alpha _4^V\alpha _2^L)\hat v{\theta ^2} + (\alpha _2^L\alpha _3^V - \alpha _5^L)\theta {\hat v^2} + \alpha _2^L{\hat v^3}\ ,
\\ 
{\varsigma _1} &= \left( {\alpha _1^L\alpha _5^V - \alpha _5^L\alpha _1^V + \alpha _4^V\alpha _2^L - \alpha _2^V\alpha _4^L} \right){\theta ^3} + \left( {\alpha _1^L\alpha _3^V + 3\alpha _2^L - \alpha _4^L} \right){\hat v ^2}\theta + (\hat v \alpha _1^L - 3\alpha _2^L){\hat v ^2} \\ &+ 2(\alpha _5^L - \alpha _2^L\alpha _3^V)\hat v \theta  + (\alpha _1^L\alpha _4^V - \alpha _1^V\alpha _4^L + 2\alpha _2^L\alpha _3^V - 2\alpha _5^L)\hat v {\theta ^2}  + (\alpha _5^L\alpha _1^V - \alpha _4^V\alpha _2^L){\theta ^2}\ ,  
\\
{\varsigma _2} &= (\alpha _1^L\alpha _4^V - \alpha _1^V\alpha _4^L + \alpha _2^L\alpha _3^V - \alpha _2^V\alpha _3^L - \alpha _5^L + \alpha _5^V){\theta ^3} + ({\hat v ^2} - 3\hat v \alpha _1^L + 3\alpha _2^L)\hat v  \\ &+ (2\alpha _1^L\alpha _3^V - \alpha _1^V\alpha _3^L + 3\alpha _2^L - 2\alpha _4^L + \alpha _4^V)\hat v {\theta ^2} + (\alpha _1^V\alpha _4^L - \alpha _1^L\alpha _4^V - 2\alpha _2^L\alpha _3^V + 2\alpha _5^L){\theta ^2} \\ &+ (3\alpha _1^L - \alpha _3^L + \alpha _3^V){\hat v ^2}\theta  + 2(\alpha _4^L - \alpha _1^L\alpha _3^V - 3\alpha _2^L)\hat v \theta  + (\alpha _2^L\alpha _3^V - \alpha _5^L)\theta\ ,    
\\ 
{\varsigma _3} &= (\alpha _1^L\alpha _3^V - \alpha _1^V\alpha _3^L + \alpha _2^L - \alpha _2^V - \alpha _4^L + \alpha _4^V){\theta ^3} + (3\alpha _1^L - \alpha _1^V - 2\alpha _3^L + 2\alpha _3^V)\hat v{\theta ^2} \\ &+ (\alpha _1^V\alpha _3^L - 2\alpha _1^L\alpha _3^V - \alpha _4^V - 3\alpha _2^L + 2\alpha _4^L){\theta ^2} + ( - 6\alpha _1^L + 2\alpha _3^L - 2\alpha _3^V)\hat v\theta  \\ &+ (2{\hat v^2} + \alpha _1^L\alpha _3^V + 3\alpha _2^L - \alpha _4^L)\theta  - 3{\hat v^2} + 3\hat v\alpha _1^L - \alpha _2^L\ , 
\\
{\varsigma _4} &= [(\alpha _1^L - \alpha _1^V - \alpha _3^L + \alpha _3^V){\theta ^2} + (\hat v - 2\alpha _1^L + \alpha _3^L - \alpha _3^V)\theta  - 3\hat v + \alpha _1^L](\theta  - 1)\ ,
\\
{\varsigma _5} &=  - { ( {\theta  - 1}  )^2} \ .
\end{aligned}
\end{equation}
Here, parameters $\alpha_i(i= 1 \ldots 5)$ are computed via the following expressions using the liquid and vapor co-volume and energy parameters:
\begin{equation}
\begin{aligned}
{\alpha _1} &= b({\delta _1} + {\delta _2}) - a/RT \ ,
\\
{\alpha _2} &= b\left( {b{\delta _1}{\delta _2} + a/RT} \right)\ ,  
\\
{\alpha _3} &= b\left( {{\delta _1}{\delta _2} - 1} \right),
\\
{\alpha _4} &= {b^2}\left( {{\delta _1} + {\delta _2} - {\delta _1}{\delta _2}} \right)\ ,
\\
{\alpha _5} &=  - {b^3}{\delta _1}{\delta _2}\ . 
\end{aligned}
\end{equation}
Since there is no analytical solution, \eqref{eq:quinticvL} has to be solved by iterative methods to obtain $v_L$. We use a Newton method with a starting point very close to the co-volume of the mixture in this study. Afterwards, the vapor's specific volume is obtained through the volume constraint \mbox{${v_V} = \left[ {\hat v - (1 - \theta ){v_L}} \right]/\theta $}.

In Step 4, we update the principal variables via their definitions in the first iteration (corresponding to a successive substitution iteration (SSI)) or evaluate Jacobian matrix and update the reduced principal variables for the Newton-Raphson iteration (NRI). In the case of the SSI, the new values of the reduced principal variables are obtained as the difference between the $h$ values of the vapor and liquid phase calculated via \eqref{eq:hcoefficients}. In the case of the NRI, first the error functions 
\begin{equation}\label{errorfun}
{e_\alpha } = h_\alpha ^V - h_\alpha ^L - h_\alpha ^\Delta \mathrm{~~ , for~~ } \alpha= 1 \ldots m+2 \ 
\end{equation}
and the associated Jacobian matrix
\begin{equation}\label{eq:jacobian}
{J_{\alpha \beta }} = \frac{{\partial {e_\alpha }}}{{\partial h_\beta ^\Delta }} = \frac{{\partial h_\alpha ^V}}{{\partial h_\beta ^\Delta }} - \frac{{\partial h_\alpha ^L}}{{\partial h_\beta ^\Delta }} - {\delta _{\alpha \beta }}\mathrm{~~ , for~~ } \alpha,\beta  = 1 \ldots m + 2 \ 
\end{equation}
are calculated, in which $\delta_{\alpha \beta}$ is the Kronecker delta function. Next, the resulting set of linear equations
$J\Delta {{\vec h}^\Delta } = \vec e,$
can be solved by using the Gauss elimination method with partial pivoting to compute $\Delta {\vec h^\Delta } = \vec h_{new}^\Delta  - \vec h_{old}^\Delta $ and the new values of the reduced principal variables  
\begin{equation}
\vec h_{new}^\Delta  = \vec h_{old}^\Delta  + \Delta {\vec h^\Delta } \ .
\end{equation}
In order to find the analytical expressions of the entries of the Jacobian matrix (\ref{eq:jacobian}), we used the classical $m+2$ reduced parameters including  $q_k(k= 1 \ldots m)$, $b$, and $\theta$ as the helping variables in the derivative chain rule for the required partial derivatives
\begin{equation}\label{app1}
\frac{{\partial h_\alpha ^j}}{{\partial h_\beta ^\Delta }} = \sum\limits_{k = 1}^m {\left( {\frac{{\partial h_\alpha ^j}}{{\partial q_k^j}} + \frac{{\partial h_\alpha ^j}}{{\partial {v^j}}}\frac{{\partial {v^j}}}{{\partial q_k^j}}} \right)\frac{{\partial q_k^j}}{{\partial h_\beta ^\Delta }} + \left( {\frac{{\partial h_\alpha ^j}}{{\partial {b^j}}} + \frac{{\partial h_\alpha ^j}}{{\partial {v^j}}}\frac{{\partial {v^j}}}{{\partial {b^j}}}} \right)\frac{{\partial {b^j}}}{{\partial h_\beta ^\Delta }} + \frac{{\partial h_\alpha ^j}}{{\partial {v^j}}}\frac{{\partial {v^j}}}{{\partial \theta }}\frac{{\partial \theta }}{{\partial h_\beta ^\Delta }}}\mathrm{~~ , for~~ } j = L,V \ .
\end{equation}
The required partial derivatives of the coefficients $h$  are obtained via \eqref{eq:hcoefficients}. The derivatives with respect to the reduced variable $q_k$ are
\begin{equation}
\begin{aligned}
\partial {h_\alpha }/\partial {q_k} =& 2{\delta _{\alpha k}}{\lambda _k}{q_k}\ln [(v + {\delta _1}b)/(v + {\delta _2}b)]/[({\delta _1} - {\delta _2})bRT] \mathrm{\ , for \, } \alpha= 1 \ldots m
\\
\partial {h_{m+1} }/\partial {q_k} =& 2{\lambda _k}{q_k}\left\{ {\begin{array}{*{20}{c}} {vb/\left[ {(v + {\delta _1}b)(v + {\delta _2}b)} \right] - }  \\  {\ln \left[ {(v + {\delta _1}b)/(v + {\delta _2}b)} \right]/({\delta _1} - {\delta _2})}  \\ \end{array}} \right\}/\left( {RT{b^2}} \right)\ , 
\\
\partial {h_{m+2} }/\partial {q_k} =& 0\ .
\end{aligned}
\end{equation}
In addition, the derivatives with respect to the co-volume of the phase are 
\begin{equation}
\begin{aligned}
\partial {h_\alpha }/\partial b &= 2{\lambda _\alpha }{q_\alpha }\left\{ {\begin{array}{*{20}{c}}   {vb/\left[ {(v + {\delta _1}b)(v + {\delta _2}b)} \right] - }  \\ {\ln \left[ {(v + {\delta _1}b)/(v + {\delta _2}b)} \right]/({\delta _1} - {\delta _2})}  \\ \end{array}} \right\}/(RT{b^2}) \mathrm{\ , for \, } \alpha= 1 \ldots m
\\
\partial {h_{m+1} }/\partial b &= av\left\{ {\begin{array}{*{20}{c}} {2\ln \left[ {(v + {\delta _1}b)/(v + {\delta _2}b)} \right]/[bv({\delta _1} - {\delta _2})] - }  \\ {[4\,{\delta _1}{\delta _2}\,{b^2} + 3\,vb({\delta _1} + {\delta _2}) + 2\,{v^2}]/{{\left[ {(v + {\delta _1}b)(v + {\delta _2}b)} \right]}^2}}  \\ \end{array}} \right\}/(RT{b^2})   - 1/(v - b), 
\\
\partial {h_{m+2} }/\partial b &= {{-1/(v - b)}}\ ,
\end{aligned}  
\end{equation}
where $a$ is computed via \eqref{eq:areducedspace} as a function of reduced parameters
The derivatives respect to the specific volume of the phase are
\begin{equation}
\begin{aligned}
\partial {h_\alpha }/\partial v &=  - 2{\lambda _\alpha }{q_\alpha }/[RT(v + {\delta _1}b)(v + {\delta _2}b)] \mathrm{\ , for \, } \alpha= 1 \ldots m
\\
\partial {h_{m+1} }/\partial v &= a[2\,b{\delta _1}{\delta _2} + v({\delta _1} + {\delta _2})]/[RT\,{{(v + {\delta _1}b)}^2}{{(v + {\delta _2}b)}^2}] + 1/(v - b)\ ,
\\
\partial {h_{m+2} }/\partial v &= {{1/(v - b)}}\ .
\end{aligned}  
\end{equation}

Next, the partial derivatives of the specific volume in \eqref{app1} are obtained through the implicit function theorem. For PT-flashes, we  directly utilize the general cubic EoS (\ref{GCEoS}) for each phase as follows:
\begin{equation}
\partial {v^j}/\partial q_k^j =  - {(\partial p/\partial {q_k})^j}/{(\partial p/\partial v)^j}\mathrm{~~ , for~~ } j = L,V
\end{equation}
with ${\partial p/\partial v}$ from \eqref{eq:dpdv}. By using the relationship between the $a$ and $q_k$, we can compute $\partial p/\partial {q_k}$ as 
\begin{equation}
{\partial p/\partial {q_k} =  - 2{\lambda _k}{q_k}/[(v + {\delta _1}b)(v + {\delta _2}b)]}.
\end{equation}
Moreover, the derivatives with respect to the co-volume of the mixture are  
\begin{equation}
\partial {v^j}/\partial {b^j} =  - {(\partial p/\partial b)^j}/{(\partial p/\partial v)^j}\mathrm{~~ , for~~ } j = L,V
\end{equation}
with
\begin{equation}
\partial p/\partial b = RT/{(v - b)^2} + a[2\,{\delta _1}{\delta _2}b + v({\delta _1} + {\delta _2})]/{[(v + {\delta _1}b)(v + {\delta _2}b)]^2}.
\end{equation}
It is obvious that, in PT-flashes, the partial derivatives of the specific volumes with respect to the vapor mole fraction are zero, i.e.
\begin{equation}
\partial {v^j}/\partial \theta  = 0\mathrm{~~ , for~~ } j = L,V \ .
\end{equation}

For utilizing the implicit function theorem for the VT-flashes, we define the function $f \equiv {p^L} - {p^V}$ and then compute the required derivatives as
\begin{eqnarray}
\partial {v^j}/\partial q_k^j =&  - (\partial f/\partial q_k^j)/(\partial f/\partial {v^j})&\mathrm{~~ , for~~ } j = L,V \ ,
\\
\partial {v^j}/\partial {b^j} =&  - (\partial f/\partial {b^j})/(\partial f/\partial {v^j})&\mathrm{~~ , for~~ } j = L,V \ ,
\end{eqnarray}
where partial derivatives of $f$ can be computed using the chain rule. For instance, when $j=L$, we obtain
\begin{eqnarray}
\partial f/\partial {v^L} =& {\left( {\partial p/\partial v} \right)^L} - {\left( {\partial p/\partial v} \right)^V}(\partial {v^V}/\partial {v^L})\ , 
\\ 
\partial f/\partial q_k^L =& {\left( {\partial p/\partial {q_k}} \right)^L} - {\left( {\partial p/\partial {q_k}} \right)^V}(\partial q_k^V/\partial q_k^L) \ , 
\\ 
\partial f/\partial {b^L} =& {\left( {\partial p/\partial b} \right)^L} - {\left( {\partial p/\partial b} \right)^V}(\partial {b^V}/\partial {b^L}) \ ,
\end{eqnarray}
along with $\partial {v^V}/\partial {v^L} = \partial q_k^V/\partial q_k^L = \partial {b^V}/\partial {b^L} = (\theta-1 )/\theta $. Subsequently, the partial derivatives of specific volumes respect to the vapor mole fraction are computed through
\begin{equation}
\partial {v^j}/\partial \theta  =  - {(\partial f/\partial \theta )^j}/(\partial f/\partial {v^j})\mathrm{~~ , for~~ } j = L,V \ ,
\end{equation}
where  ${\partial f/\partial \theta }$ for the liquid and vapor phases are  
\begin{eqnarray}
{\left( {\partial f/\partial \theta } \right)^L} =& [\sum\limits_{k = 1}^m {(q_k^L - q_k^V)(\partial f/\partial q_k^V) + } ({b^L} - {b^V})(\partial f/\partial {b^V}) + ({v^L} - {v^V})(\partial f/\partial {v^V})]/\theta\ ,
\\
{\left( {\partial f/\partial \theta } \right)^V} =& [\sum\limits_{k = 1}^m {(q_k^V - q_k^L)(\partial f/\partial q_k^L) + } ({b^V} - {b^L})(\partial f/\partial {b^L}) + ({v^V} - {v^L})(\partial f/\partial {v^L})]/(1 - \theta )\ .
\end{eqnarray}

Finally, partial derivatives of the reduced parameters $q_k(k= 1 \ldots m)$ as well as $b$ with respect to principal variables $h_\beta ^\Delta (\beta= 1 \ldots m+2)$ can be obtained via their definitions: for all $m$ reduced parameters
\begin{equation}
\partial {q_k}/\partial h_\beta ^\Delta  = \sum\limits_{i = 1}^n {\left( {\partial {z_i}/\partial h_\beta ^\Delta } \right){{\hat s}_{ki}}} \ ,
\end{equation}
and for the co-volume parameter
\begin{equation}
\partial b/\partial h_\beta ^\Delta  = \sum\limits_{i = 1}^n {\left( {\partial {z_i}/\partial h_\beta ^\Delta } \right){{\hat b}_i}} \ .
\end{equation}
In both equations, we need the derivatives of the phase mole fractions $z_i$, which is equal to $x_i$ and $y_i$ for liquid and vapor phases, with respect to the principal variables. Using the Rachford-Rice equation and the definition of the equilibrium ratio, we obtain
\begin{equation}
\partial {x_i}/\partial h_\beta ^\Delta  = {d_i}\left[ {\theta \partial {K_i}/\partial h_\beta ^\Delta  + ({K_i} - 1)\partial \theta /\partial h_\beta ^\Delta } \right] \ ,
\end{equation}
and
\begin{equation}
\partial {y_i}/\partial h_\beta ^\Delta  = {d_i}\left[ {(1 - \theta )\partial {K_i}/\partial h_\beta ^\Delta  + {K_i}({K_i} - 1)\partial \theta /\partial h_\beta ^\Delta } \right] \ ,
\end{equation}
where ${d_i} =  - {z_i}/{[1 + \theta ({K_i} - 1)]^2}$. The partial derivative with respect to principal variables is expressed as follows for all K-values: 
\begin{equation}
\begin{aligned}
\partial {K_i}/\partial h_\beta ^\Delta &= {K_i}{\hat s_{\beta i}} \ , \mathrm{for} \, \beta= 1 \ldots m
\\
\partial {K_i}/\partial h_{m + 1}^\Delta  &= {K_i}{\hat b_i} \ ,
\\
\partial {K_i}/\partial h_{m + 2}^\Delta  &= {K_i} \ ,
\end{aligned}
\end{equation}
and for the vapor mole fraction and the index the index in the range from $\beta=1$ to $m+2$:
\begin{equation}
\partial \theta /\partial h_\beta ^\Delta  = \sum\limits_{i = 1}^n {{d_i}(\partial {K_i}/\partial h_\beta ^\Delta )} /\sum\limits_{i = 1}^n {{d_i}{{({K_i} - 1)}^2}}.
\end{equation}

In Step 5, the logarithm of equilibrium ratios are computed from the updated principal variables via Eq.~\ref{eq:Kfactors} and the following convergence criterion is checked:
\begin{equation}\label{vtflashresidual}
\left\| {\ln K_i^{new} - \ln K_i^{old}} \right\| \le {\varepsilon_k} \ .
\end{equation}
We propose and use $\varepsilon_k=10^{-2}$ for the initial SSI and $\varepsilon_k = 10^{-10}$ for NRI, but one SSI step is usually enough for the most of cases. If the solution is not converged, we jump back to step 1 with the new K-values.

\subsection{Non-isothermal flashes}

\begin{algorithm}[!tb]
\SetAlgoLined
\KwResult{Equilibrium temperature}
\vspace{2mm}
Step 0: Estimate the initial value of temperature\;
\While{ convergence criteria not met }{
Step 1: Execute one VT-flash or PT-flash according to the availability of the specific volume or pressure and the latest available temperature\;
\vspace{2mm}
Step 2: In case of UV-flash, compute the specific internal energy and $c_v$ of the mixture, or in case of HP-flash, compute the specific enthalpy and $c_p$ of the mixture\;
\vspace{2mm}
Step 3: Update the temperature and check the convergence criteria\;
\vspace{2mm}}
\caption{UV and HP flash calculations}
\label{alg:2}
\end{algorithm}

In this section, the numerical solution method for the HP- and UV-flashes are explained. The main idea is to use the most appropriate isothermal flash (that is, PT for HP and VT for UV) and iterate its input temperature in such a way that the specific internal energy (UV) or enthalpy (HP) converge to the given value. The numerical procedure is summarized in Algorithm \ref{alg:2} and explained in more detailed in the following:

In Step 0, the temperature of the mixture is estimated. To provide an initial guess at regions close to the critical point or near phase boundaries, one can estimate the temperature by considering the mixture as single-phase and then iterate the EoS for the given specific internal energy or enthalpy.

In Step 1, we perform an isothermal flash calculation using the method that most closely corresponds to the targeted non-isothermal problem, that is, we perform a PT-flash in the case of the HP problem and a VT-flash for the case of the UV-flash. It is clear that for the first iteration these iso-thermal flashes require an estimate of the K-factors (it might exist from previous data otherwise those can be estimated via Wilson's correlation as discussed above) whereas for the subsequent iterations, the previously computed values of K-factors can be used to accelerate computations.

In Step 2, the internal energy of the mixture $u^{mix}$ and its derivatives with respect to the temperature, i.e. $c_v^{mix}$, for the case of UV-flash and the specific enthalpy of the mixture $h^{mix}$ and its derivative with respect to the temperature, i.e. $c_p^{mix}$, for the case of HP-flash are computed. All derivatives are computed from the reduced variables as explained in Section \ref{sec:derivatives}.

In Step 3, the estimated temperature is updated by a Newton iteration with line search $\mathscr{L}$ in the range of $\left[ {0,1} \right]$ for the case of UV-flashes:
 \begin{equation}
{T^{new}} = T - \mathscr{L}  ( {\hat u - {u^{mix}}}  )/c_v^{mix}
\end{equation}
and for the case of HP-flashes:
 \begin{equation}
{T^{new}} = T - \mathscr{L} ( {\hat h - {h^{mix}}}  )/c_p^{mix}
\end{equation}
Using the line search $\mathscr{L}$ ensures global convergence of the algorithm and renders the temperature initial guess less important.
Subsequently, the relative error is computed, i.e. ${\varepsilon _r} = |(\hat h - {h^{mix}})/\hat h|$ or ${\varepsilon _r} = |(\hat u - {u^{mix}})/\hat u|$ for HP or UV-flashes, respectively. Steps 1 to 3 are repeated until the convergence criterion is satisfied, for the calculations presented in this paper until ${\varepsilon _r} < {10^{ - 10}}$.

\section{Numerical results}

We have developed a Fortran implementation of the proposed flash algorithms for the four discussed isothermal and non-isothermal flash calculations, and tested it for a large number of different multi-component mixtures and different cubic EoS. The selected representative cases that we will present discuss in the following use the PR EoS and the values for the critical temperatures, critical pressures and acentric factors that are listed in Table~\ref{table:critical_properties}. 

\begin{table}[!tb]
\footnotesize \centering \caption{Critical properties and acentric factors of components used in this study.}\label{table:critical_properties}
\begin{tabular}{|
>{\columncolor[HTML]{EFEFEF}}l |c|c|c|
>{\columncolor[HTML]{EFEFEF}}l |c|c|c|}
\hline
Name & \multicolumn{1}{l|}{\cellcolor[HTML]{EFEFEF}$T_c$ {[}K{]}} & \multicolumn{1}{l|}{\cellcolor[HTML]{EFEFEF}$p_c$ {[}bar{]}} & \multicolumn{1}{l|}{\cellcolor[HTML]{EFEFEF}$\omega$ {[}-{]}} & Name & \multicolumn{1}{l|}{\cellcolor[HTML]{EFEFEF}${T_c}$ {[}K{]}} & \multicolumn{1}{l|}{\cellcolor[HTML]{EFEFEF} $p_c$ {[}bar{]}} & \multicolumn{1}{l|}{\cellcolor[HTML]{EFEFEF}$\omega$ {[}-{]}} \\ \hline
C\textsubscript{1}    & 190.6   & 45.4   & 0.008   & nC\textsubscript{6}   	& 507.5   & 30.1   & 0.305   \\ \hline
C\textsubscript{2}    & 305.4   & 48.2   & 0.098   & nC\textsubscript{7}  	& 540.3   & 27.4   & 0.305   \\ \hline
C\textsubscript{3}    & 369.8   & 41.9   & 0.152   & nC\textsubscript{8}   	& 568.8   & 24.9   & 0.396   \\ \hline
nC\textsubscript{4}   & 425.2   & 37.5   & 0.193   & nC\textsubscript{10}  	& 617.9   & 21.0   & 0.484   \\ \hline
nC\textsubscript{5}   & 469.6   & 33.3   & 0.251   & nC\textsubscript{14}  	& 691.9   & 15.2   & 0.747   \\ \hline
\end{tabular}
\end{table}

\begin{table}[!tb]
\footnotesize \centering \caption{Total mixture properties of the test fluids.}\label{table:properties}
\begin{tabular}{|l|ccc|ccc| }
\hline{} \multirow{2}{*}{Properties [unit]} & \multicolumn{3}{c|}{Y8 mixture} & \multicolumn{3}{c|}{MY10 mixture} \\
\cline{2-7}{} & Point A & Point B &   Point C & Point D  &  Point E & Point F \\ \hline
Temperature [K]    		& $295.40000$ & $335.20000$ &  $375.30000$ & $509.10000$ &  $566.60000$ & $563.50000$ \\
Pressure [bar]    		& $198.10000$ & $134.50000$ &  $194.80000$ & $104.90000$ &  $75.400000$ & $32.700000$ \\
Volume [L/mol]    		& $0.0805680$ & $0.1533446$ &  $0.1273056$ & $0.2280903$ &  $0.3846589$ & $1.0596464$ \\
Enthalpy [KJ/mol]   	& -$94.704181$ & -$90.636841$ &  -$87.895981$ & -$142.74142$ &  -$124.94053$ & -$120.73426$ \\ 
Int. Energy [KJ/mol]  	& -$96.300235$ & -$92.699326$ &  -$90.375895$ & -$145.13409$ &  -$127.84086$ & -$124.19930$ \\ 
\hline
\end{tabular}
\end{table}

\begin{table}[!tb]
\footnotesize \centering \caption{Molar composition of vapor and liquid in equilibrium at the states selected for the detailed analysis of the convergence of the flash algorithms.}\label{table:molefractions}
\begin{tabular}{|l|cc|cc|cc| }
\hline{} \multirow{2}{*}{Y8} & \multicolumn{2}{c|}{Point A} & \multicolumn{2}{c|}{Point B} & \multicolumn{2}{c|}{Point C} \\
\cline{2-7}{} & Liquid & Vapor &   Liquid & Vapor  &  Liquid & Vapor \\ \hline
C\textsubscript{1}    & $0.74744792$ & $0.84906008$ &  $0.47658529$ & $0.87746005$ &  $0.60400388$ & $0.81762325$ \\
C\textsubscript{2}    & $0.06057858$ & $0.05408446$ &  $0.06296756$ & $0.05530475$ &  $0.05844115$ & $0.05652908$ \\
C\textsubscript{3}    & $0.03589832$ & $0.02725004$ &  $0.05092726$ & $0.02646516$ &  $0.03965730$ & $0.03025112$ \\
nC\textsubscript{5}   & $0.06266242$ & $0.03497518$ &  $0.13974651$ & $0.02656967$ &  $0.09067889$ & $0.04396745$ \\ 
nC\textsubscript{7}   & $0.05032462$ & $0.02204618$ &  $0.13898012$ & $0.01144221$ &  $0.09260111$ & $0.03070421$ \\ 
nC\textsubscript{10}  & $0.04308814$ & $0.01258406$ &  $0.13079327$ & $0.00275817$ &  $0.11461768$ & $0.02092489$ \\
\hline
\hline{} \multirow{2}{*}{MY10} & \multicolumn{2}{c|}{Point D} & \multicolumn{2}{c|}{Point E} & \multicolumn{2}{c|}{Point F} \\
\cline{2-7}  & Liquid & Vapor &   Liquid & Vapor  &  Liquid & Vapor \\ \hline
C\textsubscript{1}     & $0.32277170$ & $0.65714256$ &  $0.27245022$ & $0.42483512$ &  $0.07783597$ & $0.38155198$\\
C\textsubscript{2}    & $0.02889804$ & $0.04243037$ &  $0.02539431$ & $0.03444446$ &  $0.00953245$ & $0.03237280$\\
C\textsubscript{3}    & $0.03944780$ & $0.04622895$ &  $0.03581565$ & $0.04403788$  & $0.01633421$ & $0.04274357$\\
nC\textsubscript{4}   & $0.06033169$ & $0.05625849$ &  $0.05673424$ & $0.06315144$  & $0.03147630$ & $0.06330675$\\ 
nC\textsubscript{5}   & $0.04080501$ & $0.03091922$ &  $0.03964769$ & $0.04033998$  & $0.02627885$ & $0.04159069$\\ 
nC\textsubscript{6}   & $0.03095915$ & $0.01918057$ &  $0.03106314$ & $0.02897407$  & $0.02441470$ & $0.03064750$\\ 
nC\textsubscript{7}   & $0.05206707$ & $0.02668295$ &  $0.05397352$ & $0.04616557$  & $0.05015849$ & $0.04998163$\\ 
nC\textsubscript{8}   & $0.05247517$ & $0.02207942$ &  $0.05603950$ & $0.04417191$  & $0.06088237$ & $0.04873841$\\ 
nC\textsubscript{10}  & $0.31843114$ & $0.09209178$ &  $0.36069877$ & $0.24142602$  & $0.52938744$ & $0.27340711$\\ 
nC\textsubscript{14}  & $0.05381324$ & $0.00698568$ &  $0.06818296$ & $0.03245354$  & $0.17369922$ & $0.03565955$\\ 
\hline
\end{tabular}
\end{table}

\subsection{Convergence behavior and robustness}

Two mixtures with specified compositions including a synthetic condensate gas and synthetic oil are selected in order to validate and evaluate the performance of proposed flash algorithms.
%
The first fluid is the Y8 mixture introduced by \citet{yarborough1972vapor}. It is a six-component synthetic gas condensate of normal alkanes including $80.97$ C\textsubscript{1}, $5.66$ C\textsubscript{2}, $3.06$ C\textsubscript{3}, $4.57$ nC\textsubscript{5}, $3.30$ nC\textsubscript{7}, and $2.44$ nC\textsubscript{10} mole percents with a zero binary interaction matrix. With our reduction method, the latter results in only three governing equations for the reduced  variables. 
The second fluid is the MY10 mixture introduced by \citet{metcalfe1979effect}. It is a ten-component mixture with overall molar fractions of $0.35$ C\textsubscript{1}, $0.03$ C\textsubscript{2}, $0.04$ C\textsubscript{3}, $0.06$ nC\textsubscript{4}, $0.04$ nC\textsubscript{5}, $0.03$ nC\textsubscript{6}, $0.05$ nC\textsubscript{7}, $0.05$ nC\textsubscript{8}, $0.30$ nC\textsubscript{10}, and $0.05$ nC\textsubscript{14}. For this mixture,  all binary interaction coefficients are zero except those between the methane and the other components as reported by \citet{firoozabadi2002fast}. This sparse binary interaction matrix results in three non-zero eigenvalues ${\lambda _1} = 9.9574$, ${\lambda _2} = 0.0707$ and ${\lambda _3} = -0.0280$.

\begin{figure}[!htb] 
\centering
\includegraphics[width=7.5cm]{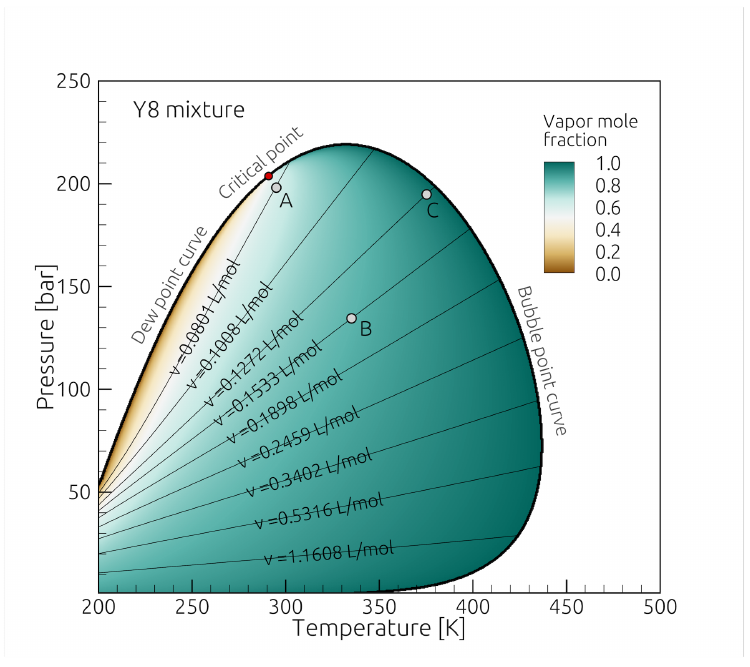} 
\includegraphics[width=7.5cm]{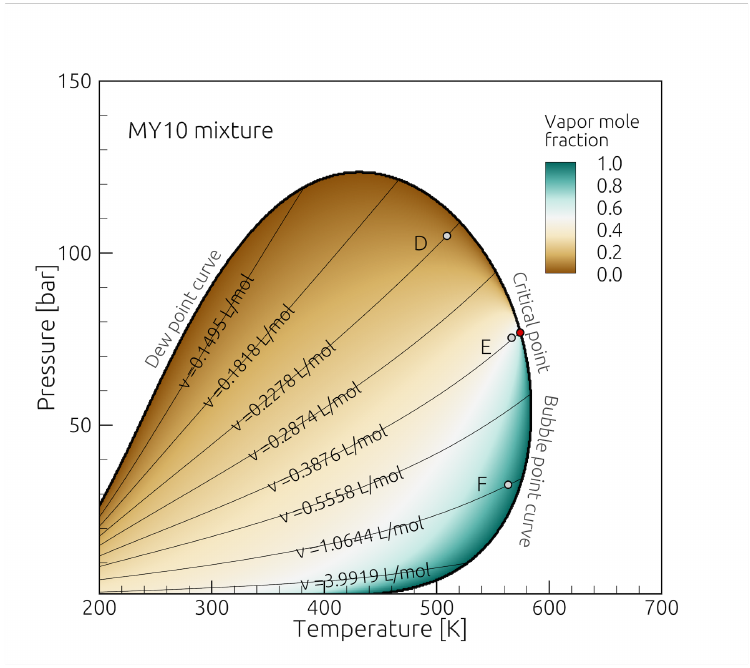} 
\caption{Phase diagrams for the Y8 and MY10 mixtures including isochores and the states selected for the detailed analysis of the convergence of the flash algorithms.}\label{fig:PTFLASH}
\end{figure}

\begin{figure}[!htb] 
\centering
\includegraphics[width=7.5cm]{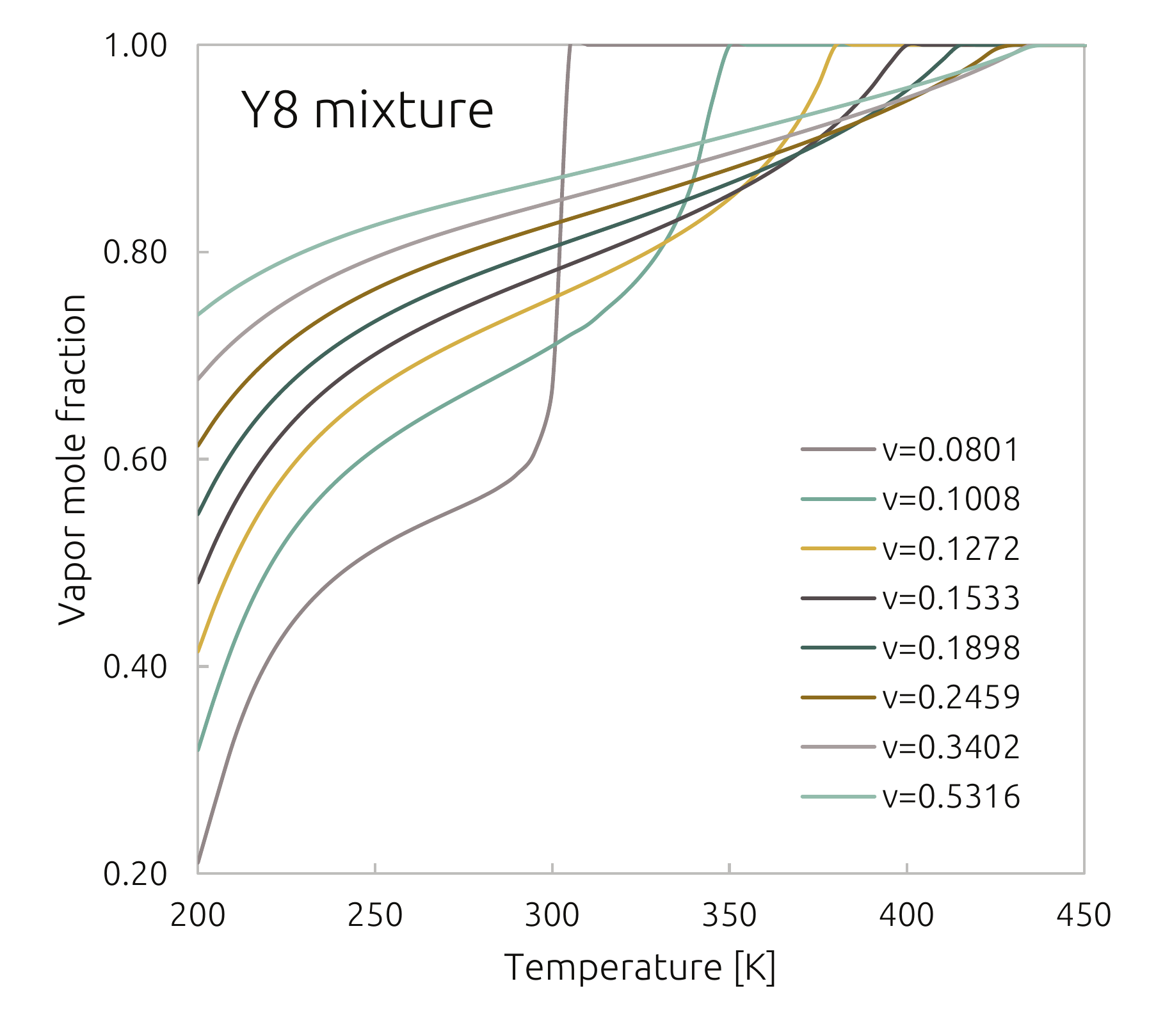} 
\includegraphics[width=7.5cm]{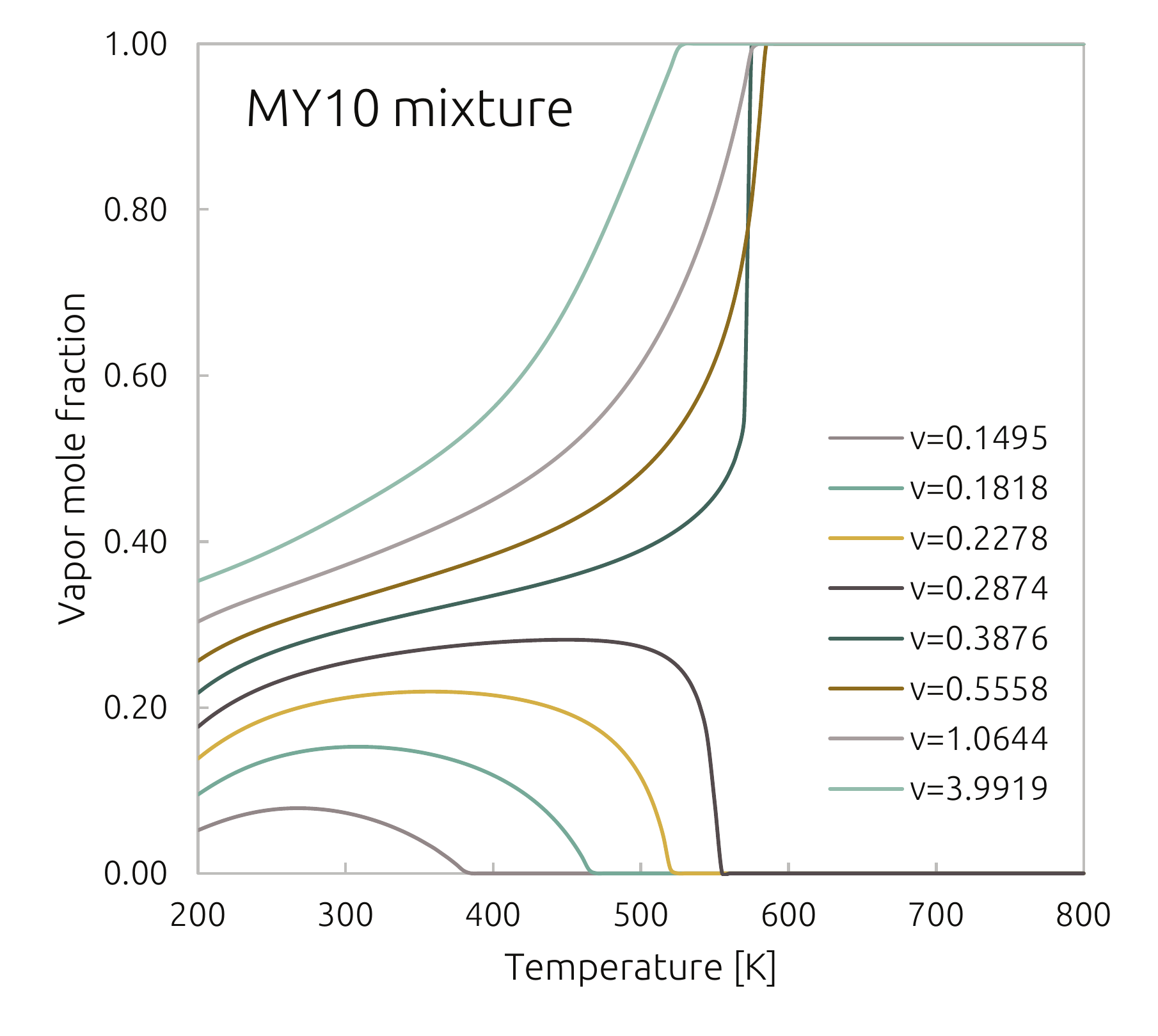} 
\caption{Vapor mole fraction curves as a function of temperature generated by the VT-flash algorithm corresponding to the lines shown in Fig.~\ref{fig:PTFLASH}.}\label{fig:VTFLASH}
\end{figure}

In Fig.~\ref{fig:PTFLASH}, the phase diagrams for the two mixtures with contours of vapor mole fractions are shown. These diagrams are generated without any convergence problems of the blind PT-flashes over a fine Cartesian grid with $800 \times 800$ nodes for the pressure - temperature range shown in the figure. This very fine grid is selected in order to check the applicability of the PT-flash algorithm at many different conditions very quite close to the phase boundaries and the critical point,  where other methods may converge either very slowly or not at all.

Next, VT-flashes have been conducted along isochores drawn on the phase diagram.
Selected isochoric lines are drawn in Fig.~\ref{fig:PTFLASH} to show the pressure evolution during constant volume heating or cooling of a typical condensate gas or synthetic oil. The corresponding vapor mole fraction curves  shown in Fig.~\ref{fig:VTFLASH}  were computed using the proposed VT-flash algorithm with the starting temperature of $200$ [K] and a step size of $1$ [K] up-to the saturation point.  The retrograde behavior of MY10 for specific volumes lower than or equal to $0.2874$ [L/mol] is an interesting phenomena: by increasing the temperature from $200$ [K] the vapor mole fraction first increases with temperature and then decreases to zero. We verified that the results agree with our previous implementation of Michelson's methods \cite{matheis2018multi} and that the results of the VT-flashes are the same as those obtained with the PT-flashes up to machine round-off precision.

The performance of the isothermal and non-isothermal flash algorithms is investigated for six algorithmically challenging points (A-F) marked in the phase diagrams of the mixtures, see Fig.~\ref{fig:PTFLASH}. The overall thermodynamic properties at these points are listed in Table~\ref{table:properties} and results for the molar composition of the vapor and liquid in equilibrium are shown in Table~\ref{table:molefractions}. These values are equal for all types of flash calculations.

\begin{figure}[!htb] 
\centering
\includegraphics[width=7.5cm]{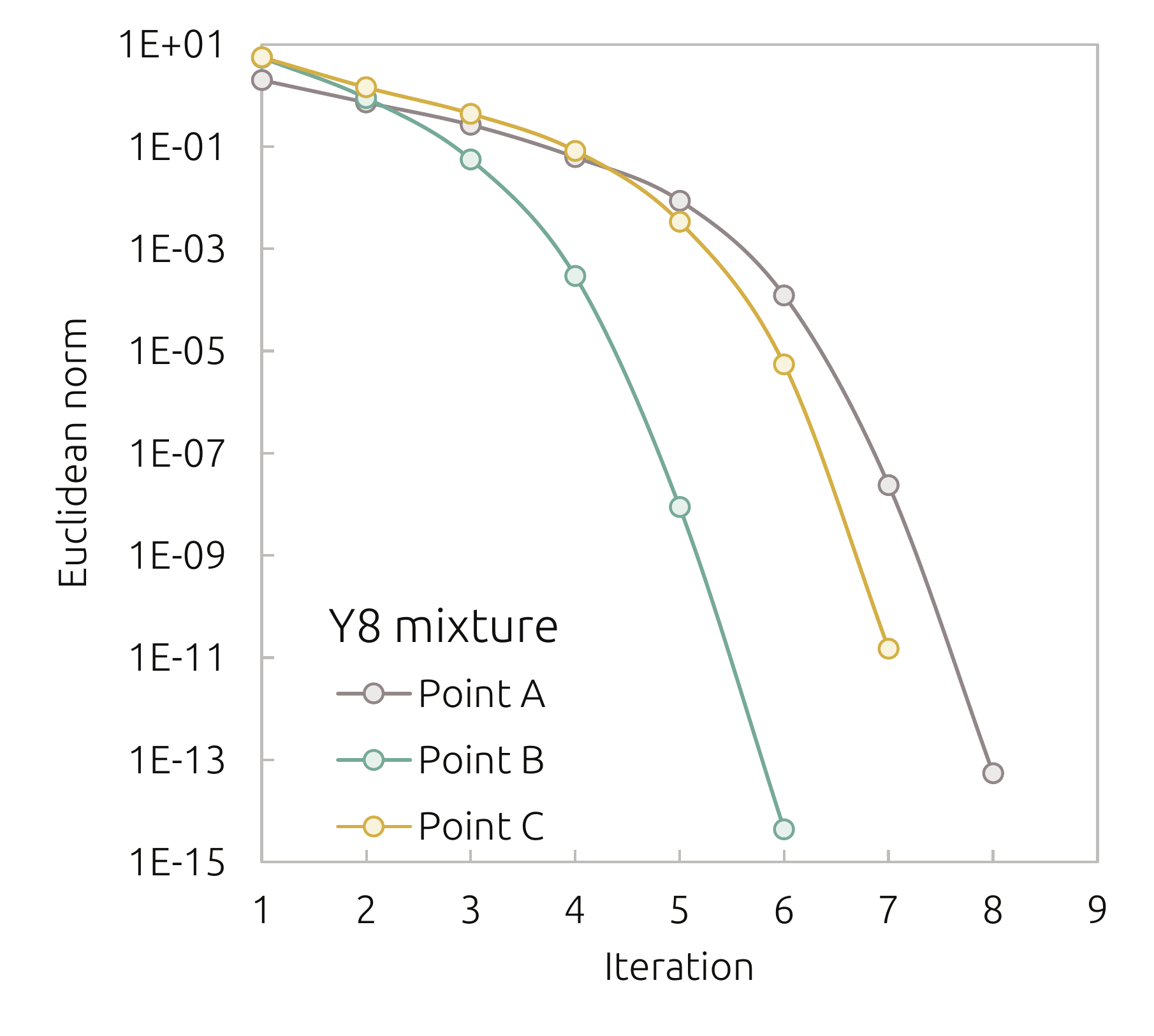} 
\includegraphics[width=7.5cm]{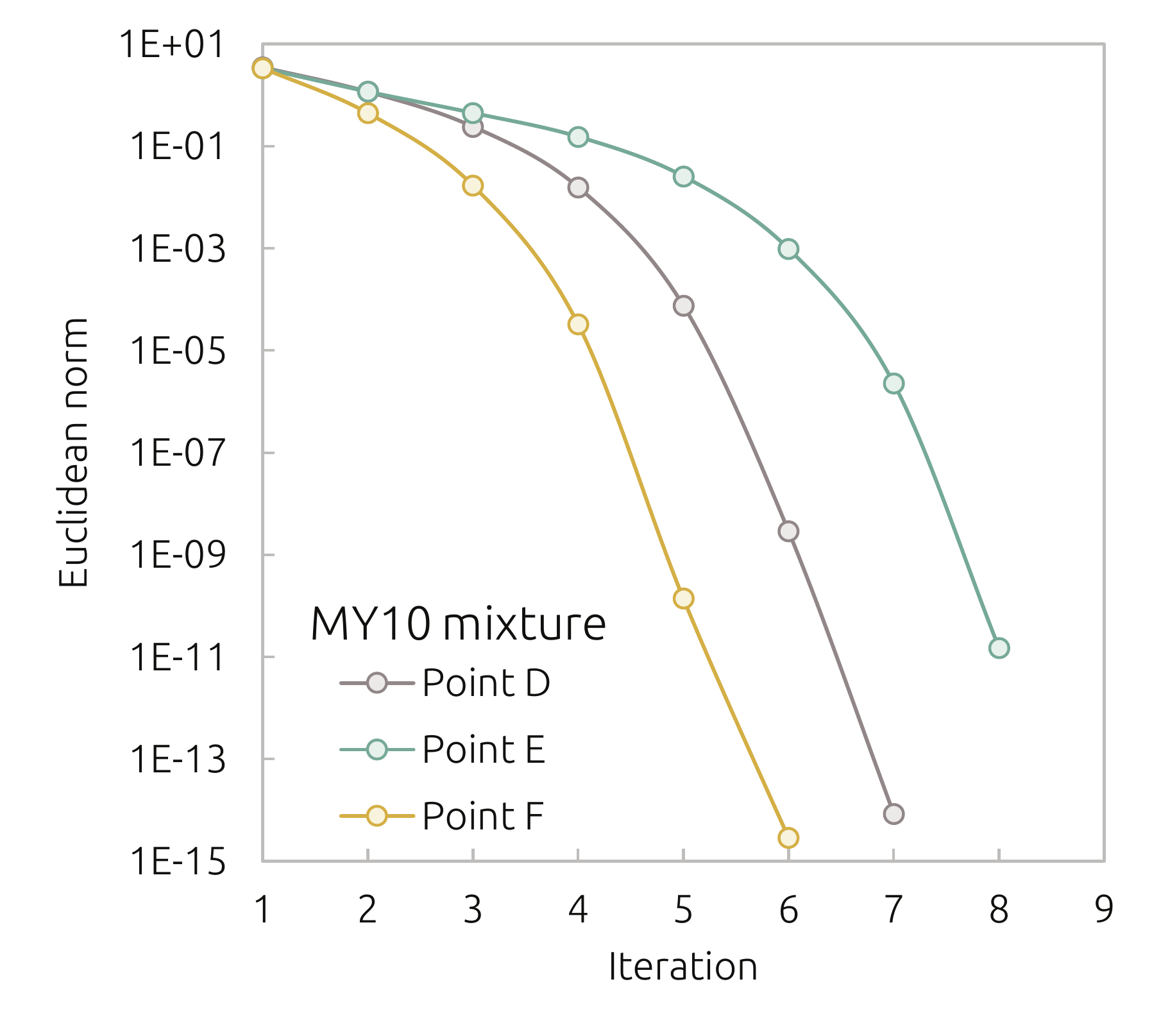} 
\caption{PT-flash convergence for the Y8 and MY10 mixtures at the points marked in Fig.~\ref{fig:PTFLASH}.}\label{fig:PTconvg}
\end{figure}

\begin{figure}[!htb] 
\centering
\includegraphics[width=7.5cm]{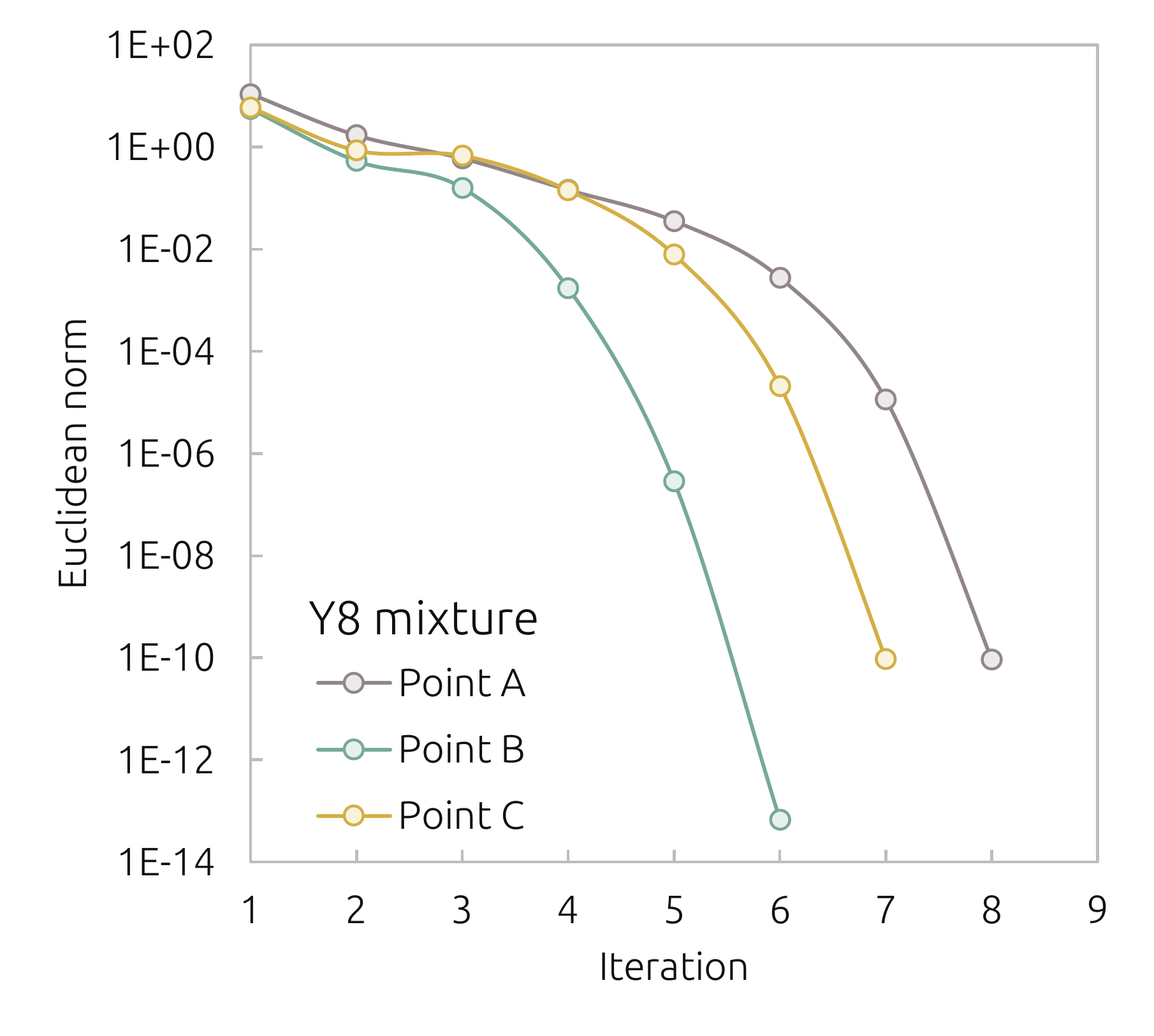} 
\includegraphics[width=7.5cm]{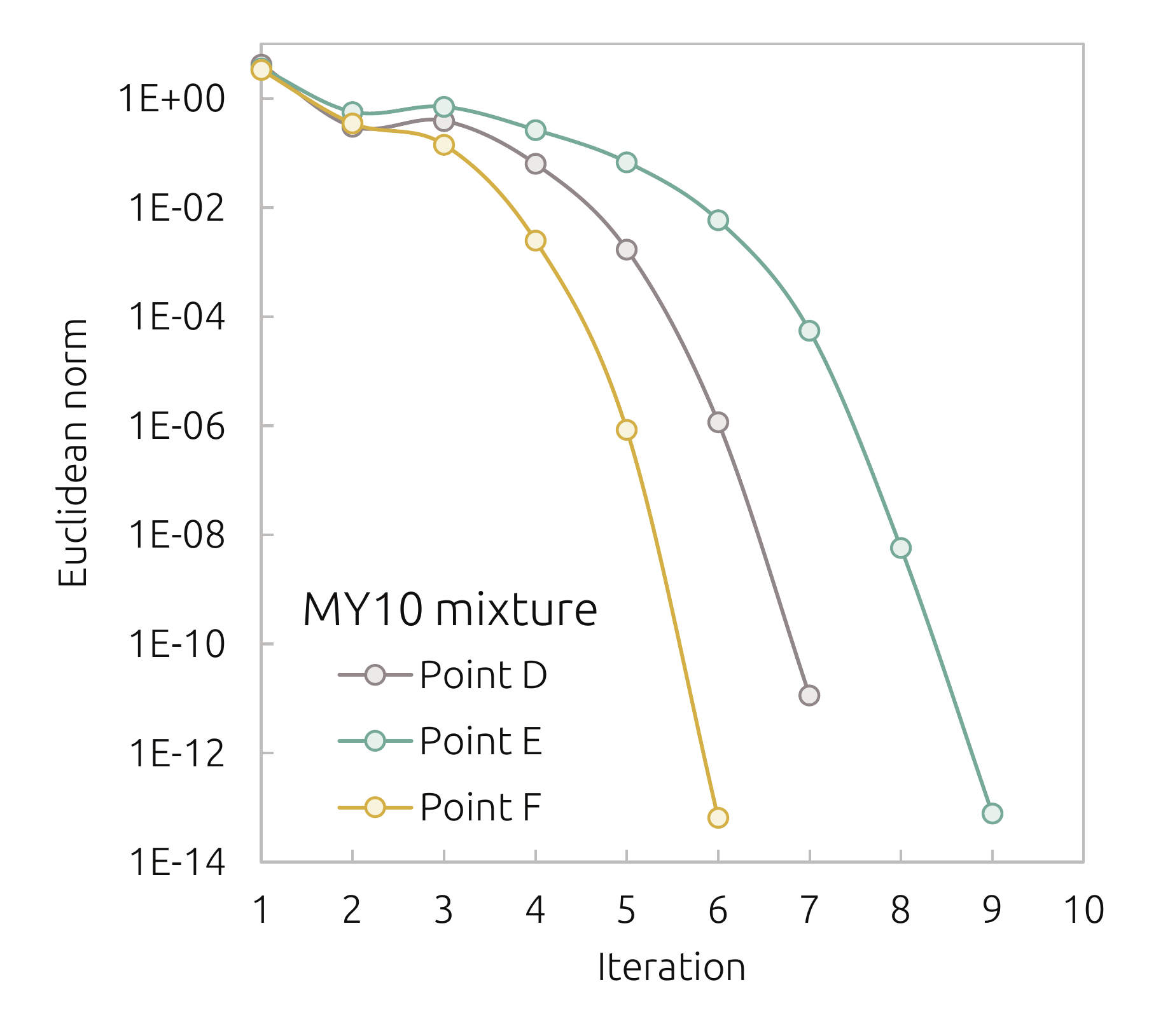} 
\caption{VT-flash convergence for the Y8 and MY10 mixtures at the points marked in Fig.~\ref{fig:PTFLASH}.}\label{fig:VTconvg}
\end{figure}

\begin{figure}[!htb] 
\centering
\includegraphics[width=7.5cm]{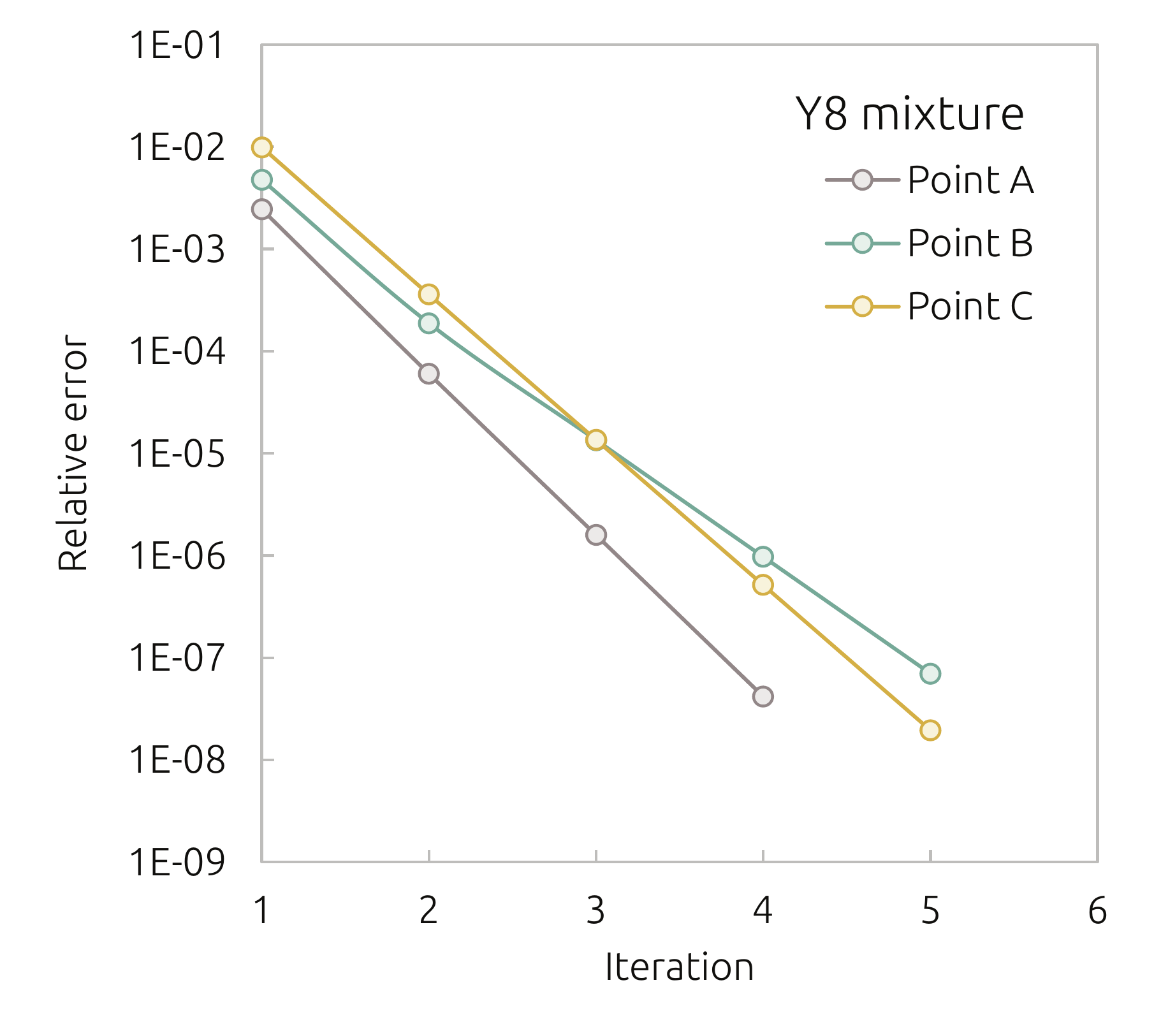} 
\includegraphics[width=7.5cm]{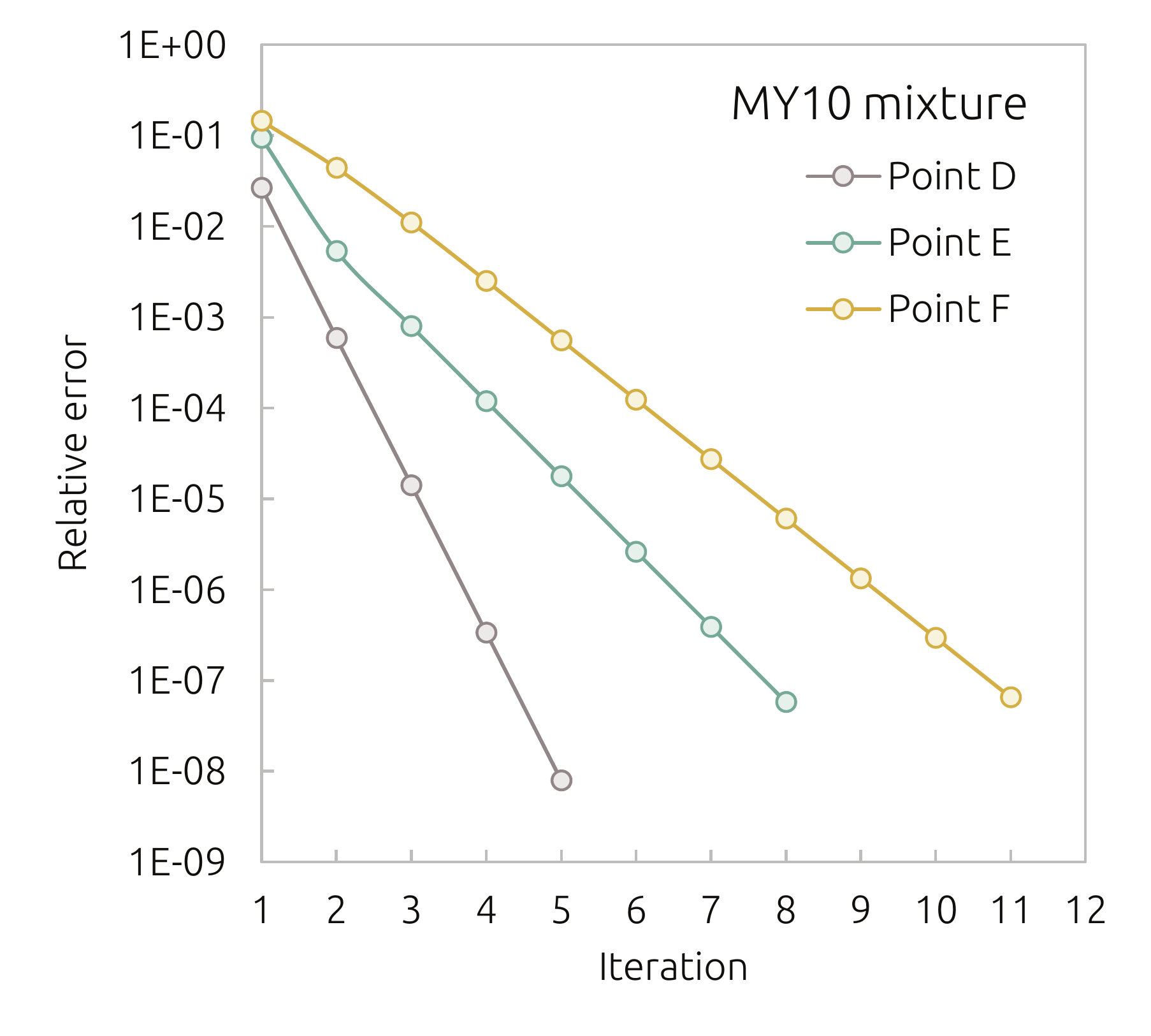} 
\caption{HP-flash convergence for the Y8 and MY10 mixtures at the points marked in Fig.~\ref{fig:PTFLASH}.}\label{fig:HPconvg}
\end{figure}

\begin{figure}[!htb] 
\centering
\includegraphics[width=7.5cm]{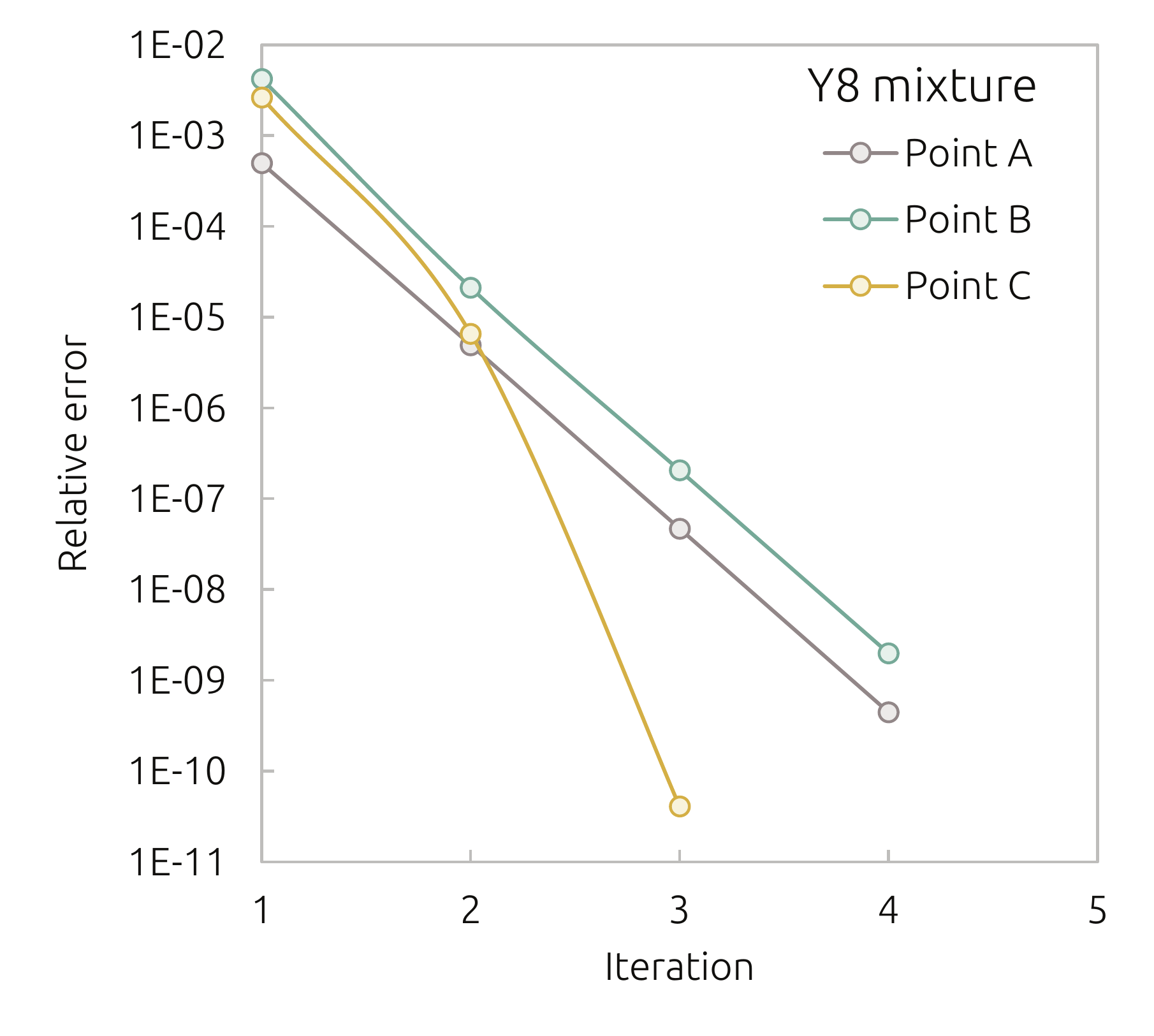} 
\includegraphics[width=7.5cm]{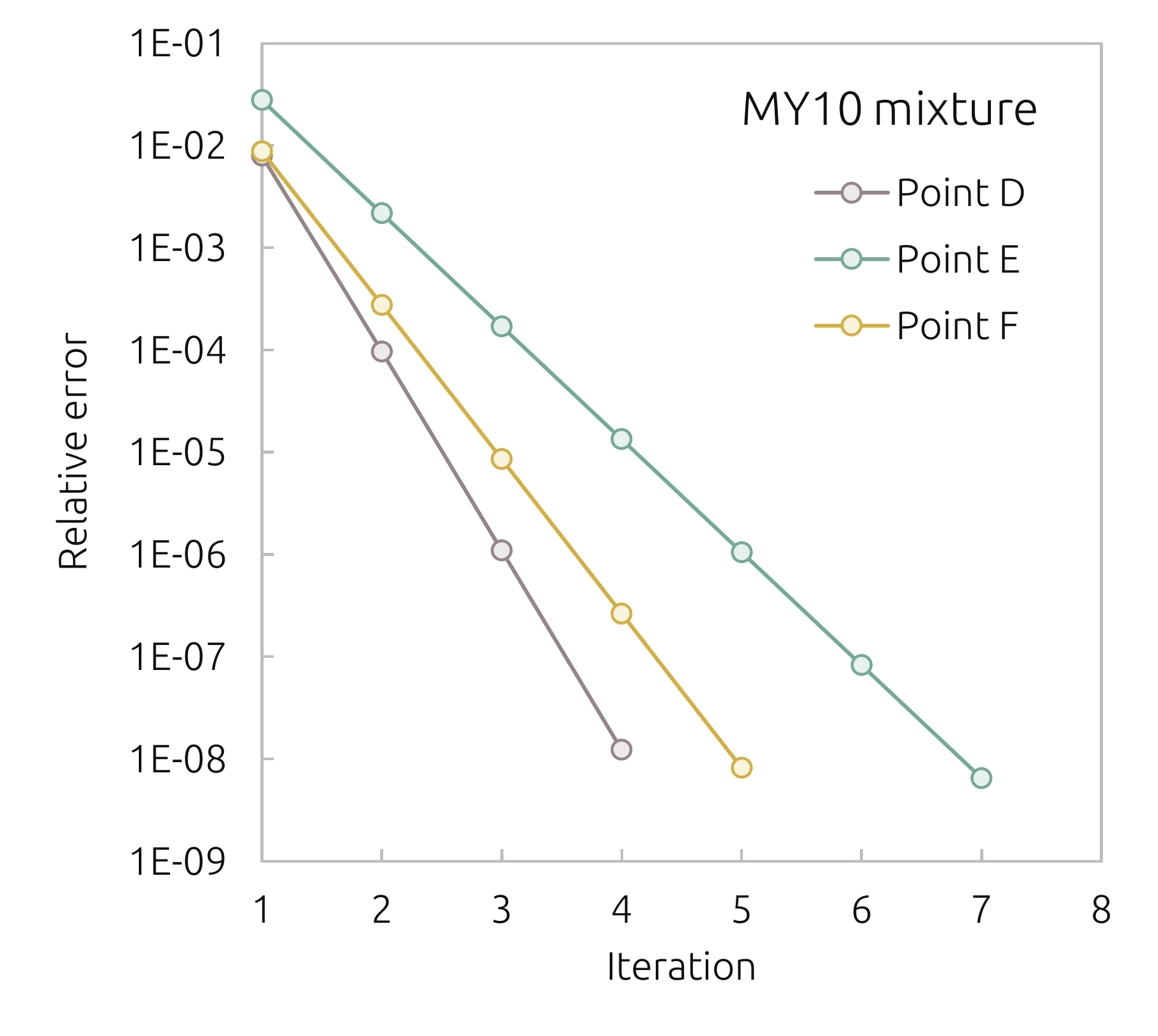} 
\caption{UV-flash convergence for the Y8 and MY10 mixtures at the points marked in Fig.~\ref{fig:PTFLASH}.}\label{fig:UVconvg}
\end{figure}

The evolution of the Euclidean residual norm for the PT-flash calculations at points A~($198.1$~[bar], $295.4$~[K]), B~($134.5$~[bar], $335.2$~[K]), and C~($194.8$~[bar], $375.3$~[K]) for the Y8 mixture, and at points D~($104.9$~[bar], $509.1$~[K]), E~($75.4$~[bar], $566.6$~[K]), and F~($32.7$~[bar], $566.6$~[K]) for the MY10 mixture are plotted in Fig.~\ref{fig:PTconvg}. For all points, one SSI has been carried out on the initial K-factors obtained from Wilson's correlation before switching to the NRI. Results indicate that PT-flash algorithm requires about $6-8$ iterations for the points at the heart of the two-phase dome, near the phase boundaries, and close to the critical point.

The convergence of the residuals of VT-flash calculations are shown in Fig.~\ref{fig:VTconvg} for the points A~($0.0806$~[L/mol], $295.4$~[K]), B~($0.1533$~[L/mol], $335.2$~[K]), and C~($0.1273$~[L/mol], $375.3$~[K]) corresponding to the marked points on the phase diagram of the Y8 mixture, and for the points D~($0.2281$~[L/mol], $509.1$~[K]), E~($0.3847$~[L/mol], $566.6$~[K]), and F~($1.0596$~[L/mol], $563.5$~[K]) corresponding to the tagged points on the phase diagram of the MY10 mixture. Initial values for the K-factors were obtained from Wilson's correlation using a pressure obtained from the state equation by the overall composition and given temperature and volumes, i.e. $p = \lbrace 188.8, 119.3, 193.2, 95.5, 74.1, 33.2 \rbrace$ [bar] for points \mbox{$\lbrace$ A to F $\rbrace$}. As for the PT-flashes, one initial SSI was executed before switching to NRI. The VT-flash results show the same excellent convergence behavior as observed for the PT-flashes, that is, both algorithms have optimum quadratic convergence and require only very few iterations, with only two iterations difference between points close to and far from the extreme conditions.

Figure~\ref{fig:HPconvg} shows the convergence of the relative errors for the blind HP-flash at points A~($-94.704$~[kJ/mol], $198.1$~[bar]), B~($-90.637$~[kJ/mol], $134.5$~[bar]), and C~($-87.896$~[kJ/mol], $194.8$~[bar]) for the the condensate gas and at points D~($-142.741$~[kJ/mol], $104.9$~[bar]), E~($-124.941$~[kJ/mol], $75.4$~[bar]), and F~($-120.734$~[kJ/mol], $32.7$~[bar]) for the synthetic oil. We initalize the iteration with a temperature of $250$ [K] for Y8 and $400$ [K] for MY10, far away from the true solution, in order to test the robustness of the non-isothermal flashes at extreme conditions. As the convergence plots show, they rapidly converge within very few iterations even with a poor inital temperature guess.

Figure~\ref{fig:UVconvg} shows results for the blind UV-flash at the points A~($-96.300$~[kJ/mol], $0.0805$~[L/mol]), B~($-92.699$~[kJ/mol], $0.1533$~[L/mol]), and C~($-90.376$~[kJ/mol], $0.1273$~[L/mol])  for Y8 and at the points D~($-145.134$~[kJ/mol], $0.2281$~[L/mol]), E~($-127.841$~[kJ/mol], $0.3847$~[L/mol]), and F~($-124.199$~[kJ/mol], $1.0596$~[L/mol]) for the MY10 mixture. As before, the initial temperature guess is $250$ [K] for the Y8 gas condensate and $400$ [K] for the MY10 oil mixture. The initial values for the pressure are the same as used for the VT-flash at these points, see above. We observe rapid convergence within 4 to 7 iterations to within a relative error of $10^{-8}$. For the most of engineering applications, it is, however, not necessary to know the temperature with such a high precision and a much larger error, say $0.1$ [K] can be tolerated. The algorithms for both non-isothermal flashes yield temperature differences of less than $0.1$ [K] in just $3$ iterations.

\subsection{Computational time}

\begin{figure}[!htb]
\centering
\includegraphics[width=9cm]{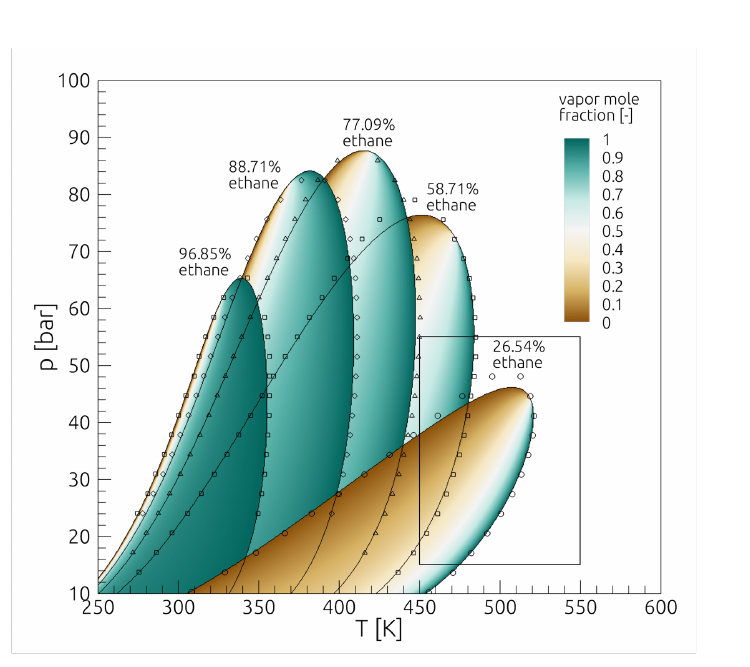} 
\caption{Phase diagram of binary mixtures of n-heptane and ethane at various molar compositions computed by the proposed algorithm. The symbols denote  experimental reference data for the dew-point and bubble-point lines \citep{Kay:1938dw,Mehra:1965jq}. The black box encloses the pressure-temperature domain that was used for measuring the computational performance of the flash algorithms.}\label{fig:EthaneHepthane}
\end{figure}

\begin{figure}[!htb] 
\centering
\includegraphics[width=7.5cm]{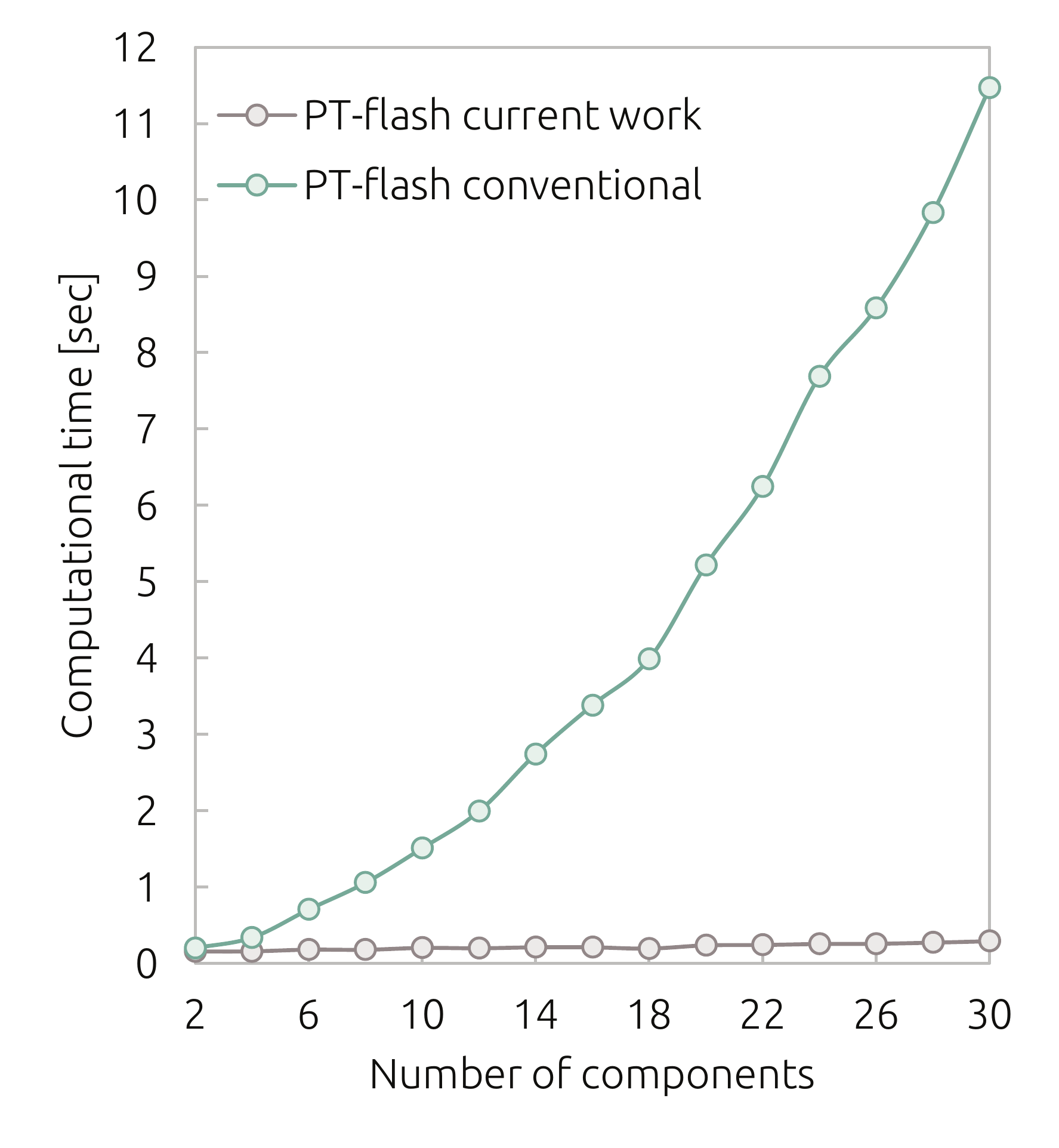} 
\includegraphics[width=7.5cm]{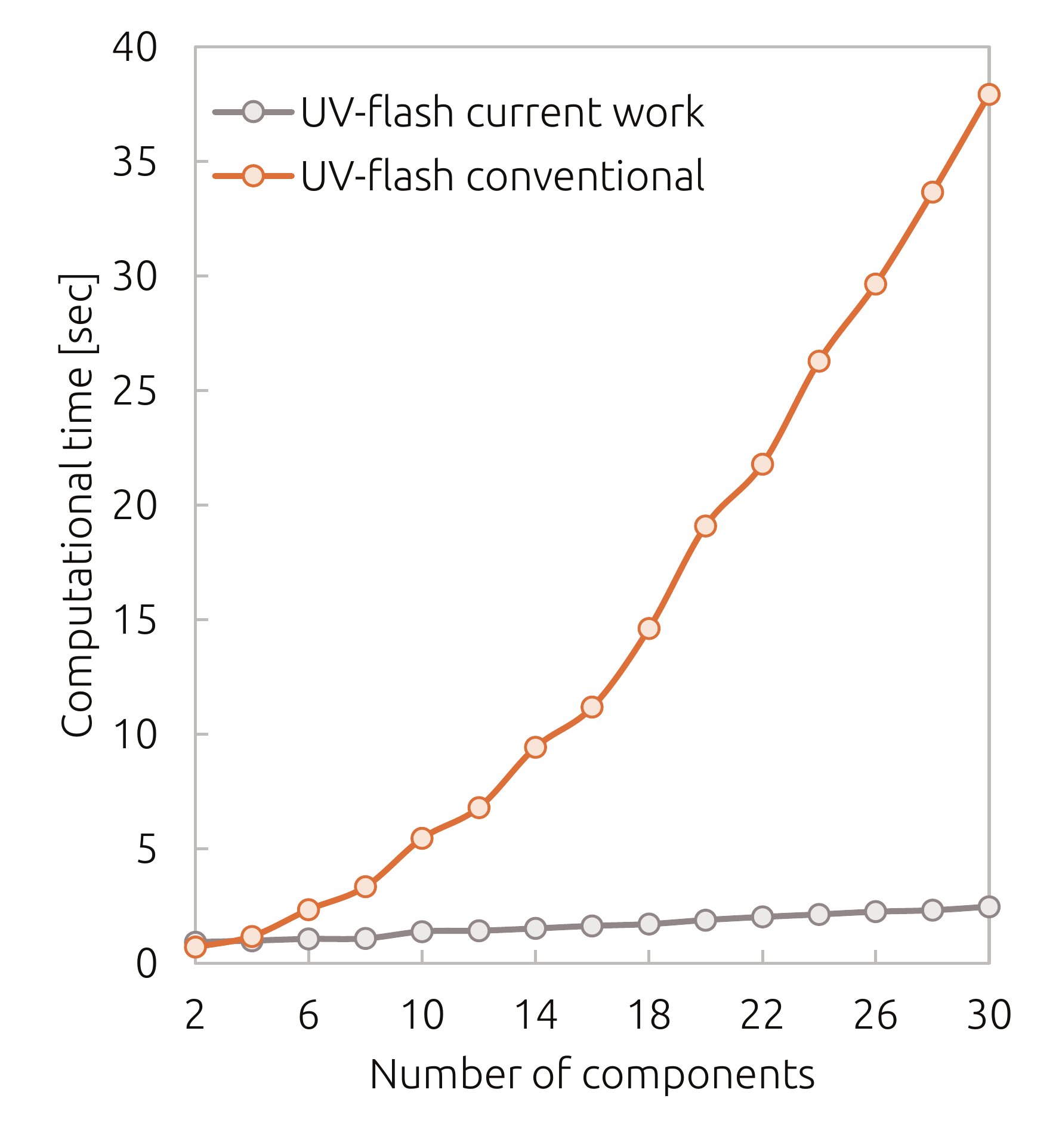} 
\caption{Computational time for PT-flashes and UV-flashes vs.\ number of mixture components. Shown is the total CPU time for $100 \times 100$ flash calculations in the part of the phase diagram highlighted in Fig.~\ref{fig:EthaneHepthane}.}\label{fig:speedup}
\end{figure}

In this section, we analyse the computational performance of the proposed flash algorithms for different mixtures and demonstrate the improved efficiency resulting from using a reduction method and direct VT-flashes instead of PT-flashes in the inner iteration loop of UV-flashes. For a fair quantitative evaluation, the computational time required for the new flash algorithms that we propose in this paper is compared with the highly optimized implementation of a conventional method that was developed by \citet{Matheis:2016td} for the large-scale turbulence-resolving CFD simulations of transcritical fuel injection. Both algorithms use a Newton method to achieve fast convergence. 
The calculations are performed on an Intel Xeon W-2123 CPU at 3.60 GHz and the Intel Fortran compiler was used.

The test fluid is a mixture of ethane and normal heptane. The phase diagrams of this binary mixture for various molar compositions are shown in  Fig.~\ref{fig:EthaneHepthane}. To study the effect of the component number in the mixture, we have added pseudo-components with properties identical to ethane and normal heptane and adjust the mole fractions in a way that the total composition remains constant. First, PT-flash calculations were carried out for a mixture with $26.54\%$ ethane and we record the total computational time $100 \times 100$ states in the pressure-temperature range that is enclosed by the black box in Fig.~\ref{fig:EthaneHepthane}. Then, the mixture internal energy and specific volume that were computed by the PT-flashes are used for executing the corresponding UV-flashes. In order to assess the performance of the proposed UV-flash at conditions that similar to what we typically encounter in CFD simulations, initial guesses of pressure and temperature were computed by adding random perturbations to the true values, that is:
\begin{subequations}
\begin{align}
\begin{split}
T_{guess} &= {T_{true}} + r \Delta T \ , 
\end{split}
\\
\begin{split}
p_{guess} &= {p_{true}} + r\Delta p \ ,
\end{split}
\end{align}
\end{subequations}
where $r$ is a random number generated in the range $[-0.5,0.5]$. The perturbation amplitudes $\Delta T$ and $\Delta p$ are set to $20$ [K] and $20$ kPa, which corresponds to the maximum change that we can expect between two subsequent time steps in CFD simulations. 

The results are shown in Fig.~\ref{fig:speedup}. The computational time for the current PT-flash algorithm is always lower than the highly optimized reference method. The difference becomes more significant as the number of components is increased, which shows the importance of reduction methods for the both iso-thermal and non-isothermal flashes. Surprisingly, we also measure a performance gain for the two-component mixture, where the number of variables is not reduced by the new method. In this case, the reduction method acts as a preconditioner and reduces the number of required iterations for the PT-flash. Furthermore, it should be noted that the computational performance the UV-flash based on the VT-flash is much less sensitive to the amplitude of the imposed pressure perturbation $\Delta p$ than the conventional method based on the PT-flash. For instance, the conventional method becomes more than five times slower for $\Delta p=400$ kPa, whereas the overall time needed for the new method remains unchanged.

\section{Discussion and Conclusions}

This work was motivated by the need for computationally efficient and very robust vapor-liquid phase-split calculations in turbulence-resolving CFD simulations of high-pressure liquid-fuel injection and reacting transcritical multiphase flows in modern energy conversion systems, such as rocket engines, gas turbines and jet engines. Such simulations require typically $10^{10}$ to $10^{16}$ flash calculations for given overall specific internal energy, volume and composition and unknown pressure, temperature, volume fractions and phase compositions (isoenergetic-isochoric flash -- UV-flash). The standard methods for such applications that we used in the past \citep{Matheis:2016td,matheis2018multi} are based on a nested PT-flash and suffer from poor conditioning near the spinodal and coexistence curves and polynomial growth of the computational cost in terms of the number of mixture components. To this end, and building upon and extending the work of \citet{mikyvska2011new} and \citet{nichita2011new}, we have developed a new multi-component reduction method for direct PT-flash and VT-flash calculations based on the formulation of phase equilibrium conditions in terms of the molar specific value of \citeauthor{mikyvska2011new}'s volume function and a corresponding adaptation of \citeauthor{nichita2011new}'s reduction method. The computational cost of solving the PT-flash and VT-flash in terms of the new reduced set of variables is almost independent of the number of components and the point in the phase diagram. The reduced-space Newton-Raphson iteration, using the exact analytical Jacobian matrix, results in optimum quadratic convergence in very few iterations. We further showed that the non-isothermal UV and HP flashes are efficiently solved through univariate residual minimization with the naturally corresponding isothermal flash (PT-flash for HP-flash and VT-flash for UV-flash) and the specific heat capacity at constant pressure (for HP-flash) or at constant volume (for UV-flash) as exact Jacobian. We have thoroughly verified the reliability and efficiency of the algorithmic implementation. The computational results show a considerable speed-up compared to conventional methods, as well as improved robustness and better convergence behavior near the spinodal and coexistence curves of multi-component mixtures, where the preconditioning by the reduction method is most effective.

\section*{Acknowledgements}
This work is funded by the Netherlands Organisation for Scientific Research (Contract No. 680.91.082).



\bibliographystyle{unsrtnat}
\bibliography{refs}

\end{document}